\documentclass[%
 reprint,
 amsmath,amssymb,
 aps,
prd,
floatfix,
]{revtex4-2}

\usepackage{graphicx}% Include figure files
\usepackage{dcolumn}% Align table columns on decimal point
\usepackage{bm}% bold math
\usepackage{slashed}% Package for Dirac slash notation
\usepackage{color}% Color text
\usepackage{textcomp}% text edit
\usepackage{enumerate}% to set enumerate labels
\usepackage{extarrows}% for additional arrows
\usepackage{comment}% to comment out
\usepackage{float}

\usepackage{tikz}
\usetikzlibrary{calc} 
\usetikzlibrary{patterns} 
\usetikzlibrary{decorations.pathreplacing} 
\usetikzlibrary{decorations.markings} 
\usetikzlibrary{decorations.pathmorphing} 
\usetikzlibrary{positioning}

\newcommand{\eq}[1]{Eq.~\eqref{#1}}

\newcommand{\eqs}[2]{Eqs.~\eqref{#1} and \eqref{#2}}
\newcommand{\fref}[1]{Fig.~\ref{#1}}

\newcommand{\sref}[1]{Sec.~\ref{#1}}

\newcommand{\dd}{\mathrm{d}}

\newcommand{\bs}{\boldsymbol}
\newcommand{\nn}{\nonumber}

\def\mnras{Mon. Not. R. Astron. Soc. }

\def\apj{Astrophys. J.}
\def\apjl{Astrophys. J. Lett.}
\def\apjs{Astrophys. J., Suppl. Ser.}
\def\prd{Phys. Rev. D}
\def\prl{Phys. Rev. Lett.}

\def\araa{Annu. Rev. Astron. Astrophys.}

\def\baas{Bull. Am. Astron. Soc.}
\def\nat{Nature}
\def\pasp{Publ. Astron. Soc. Pac.}

\begin{document}

\preprint{APS/123-QED}

\title{Excitation of gravitational wave modes by a center-of-mass velocity of the source}

\author{Alejandro Torres-Orjuela}
\affiliation{Astronomy Department, School of Physics, Peking University, 100871 Beijing, China}
\affiliation{Kavli Institute for Astronomy and Astrophysics at Peking University, 100871 Beijing, China}

\author{Xian Chen}
\email{Corresponding author: xian.chen@pku.edu.cn}
\affiliation{Astronomy Department, School of Physics, Peking University, 100871 Beijing, China}
\affiliation{Kavli Institute for Astronomy and Astrophysics at Peking University, 100871 Beijing, China}

\author{Pau Amaro Seoane}
\affiliation{Universitat Polit{\`e}cnica de Val{\`e}ncia, 46022 Val{\`e}ncia, Spain}
\affiliation{Deutsches Elektronen Synchrotron DESY, Platanenallee 6, 15738 Zeuthen, Germany}
\affiliation{Kavli Institute for Astronomy and Astrophysics at Peking University, 100871 Beijing, China}
\affiliation{Institute of Applied Mathematics, Academy of Mathematics and Systems Science, Chinese Academy of Sciences, Beijing, China}
\affiliation{Zentrum f{\"u}r Astronomie und Astrophysik, TU Berlin, Germany}

\date{\today}

\begin{abstract}
Most gravitational wave (GW) sources are moving relative to us. This motion is often closely related to the environment of the source and can thus provide crucial information about the formation of the source and its host. Recently, LIGO and Virgo detected for the first time the subdominant modes of GWs. We show that a motion of the center-of-mass of the source can affect these modes, where the effect is proportional to the velocity of the source. The effect on the GW modes in turn affects the overall frequency of the GW, thus leading to a phase shift. We study the impact of this effect on LIGO/Virgo detections and show that it is detectable for sources with high mass ratios and inclinations. This effect breaks the degeneracy between mass and Doppler shift in GW observations, and opens a new possibility of detecting the motion of a GW source even for constant velocities.
\end{abstract}

%\keywords{keywords}%Use showkeys class option if keyword

\maketitle

\section{Introduction}\label{sec:int}

Recently, the LIGO and Virgo detectors detected gravitational waves from merging binary black holes (BBHs) with a significant contribution from modes other than the dominant $(2,2)$-mode~\cite{ligo_2020a,ligo_2020b}. For the detection of the so-called subdominant modes a relatively high signal-to-noise ratio (SNR) of $\rho \gtrsim 20$ for the two events as well as significant progresses in modelling waveforms containing subdominant modes accurately have been crucial~\cite{khan_ohme_2020,khan_chatziioannou_2019,ossokine_buonanno_2020,pan_buonanno_2014,babak_taracchini_2017,varma_field_2019}. Next-generation ground-based detectors like the Einstein Telescope and Cosmic Explorer~\cite{et_2010,cosmic_explorer_2019}, as well as space-based observatories such as LISA, Tianqin and Taiji~\cite{lisa_2017,tianqin_2016,taiji_2015} could achieve much higher SNRs of $\rho>100$ in the future. A higher SNR will then allow us to study subdominant modes in more detail, demanding their modelling to high accuracy.

A feature current GW models have in common is to compute the waveform in a frame where the center-of-mass (CoM) is initially at rest~\cite{blanchet_2006,santamaria_ohme_2010,hannam_schmidt_2014,buonanno_damour_1999,SXS_2019,RIT_2019,GT_2016,blackman_field_2017}. However, the majority of astrophysical objects are constantly moving and hence the templates used by LIGO and Virgo nowadays should be considered as an approximation. This approximations is tolerable for relatively low SNR, but in the future for detections with high SNR, accurate templates should include velocity and its effects on GWs.

Two astrophysical scenarios are of particular interest concerning a CoM velocity of GW sources. The first scenario is the peculiar velocity of galaxies. Other galaxy clusters are moving relative to our Local Group with velocities ranging from about several $100\,\textrm{km\,s}^{-1}$ to around $2000\,\textrm{km\,s}^{-1}$~\cite{bahcall_1988,scrimgeour_davis_2016,colin_mohayaee_2017}. Moreover, a significant fraction of galaxies resides in rich galaxy clusters with velocity dispersions of around $1000\,\textrm{km\,s}^{-1}$~\cite{zinn_west_1984,girardi_fadda_1996,carlberg_yee_1996,springel_white_2001,ruel_bazin_2014}. Therefore, GW sources from outside of our galaxy can almost always be considered as moving with velocities of several $100\,\textrm{km\,s}^{-1}$ to a few $1000\,\textrm{km\,s}^{-1}$. The second scenario is motivated by the theoretical prediction that a population of merging BBHs may come from triple systems, with the third body being either a star~\cite{wen_2003,naoz_2016,meiron_kocsis_2017,arca-sedda_2020} or a supermassive black hole (BH) in the center of a galaxy~\cite{antonini_perets_2012,mckernan_ford_2012,addison_gracia-linares_2019,bartos_kocsis_2017,stone_metzger_2017,tagawa_haiman_2020}. The velocities for these systems can range again from a few $100\,\textrm{km\,s}^{-1}$ but go up to several percent of the speed of light~\cite{chen_li_2017,chen_han_2018,han_chen_2018}.

The decomposition of GWs in modes represents a decomposition according to their angular properties, which is conveniently described by the so-called spin-2 spherical harmonics~\cite{ruiz_alcubierre_2008,goldberg_macfarlane_1967}. The modes a GW signal contains is not only closely related to the spherical symmetry of the source but also to the multipoles of the source, thus containing detailed information about the structure and dynamics of the source~\cite{thorne_1980}. In Ref.~\cite{torres-orjuela_chen_2019} we show that the amplitude of a GW signal is velocity-dependent. This change in the amplitude can be explained by the aberration of the GW rays and a rotation of the polarization. These two effects cause that the radiation pattern seen for a moving source differs from the one seen for a source at rest and hence can affect the modes detected for the source. The effect of velocity on the modes of GWs has been studied in the literature for non-relativistic velocities parallel to the line-of-sight, for gravitational kicks and in the context of corrections to Numerical Relativity templates~\cite{gualtieri_berti_2008,boyle_2016,woodford_boyle_2019}. In this paper, we study how velocity affects the modes of GWs for general sources and without restriction on the magnitude and orientation of the velocity by focusing on the effects of aberration and polarization rotation on the GWs. We derive how this change results in an excitement of GW modes and study the detectability of this signature by LIGO and Virgo. Throughout this paper, unless otherwise indicated, we use geometrical units in which the gravitational constant and the speed of light are equal to one (i.e., $G=c=1$).

\section{Gravitational waves modes}\label{sec:gwm}

In General Relativity, GWs have two independent components, the $+$-polarization, $h_+$, and the $\times$-polarization, $h_\times$, which can be combined to the complex amplitude, $H(\theta,\phi) := h_+(\theta,\phi) - ih_\times(\theta,\phi)$. Using this complex amplitude we can decompose the GW in its spherical components, which describe the `shape' of the source, and its time/radial components, which describe the `evolution' of the source. The spherical components are described by spin-weighted spherical harmonics, $_sY^{\ell,m}(\theta,\phi)$, of spin $s = -2$~\cite{ruiz_alcubierre_2008}
\begin{equation}\label{eq:deccoma}
    H(\theta,\phi) = \sum_{\ell = 2}^\infty\sum_{m=-\ell}^\ell H^{\ell,m}\,_{-2}Y^{\ell,m}(\theta,\phi).
\end{equation}
The $H^{\ell,m}$ are denoted as the $(\ell,m)$-modes of the GW and are only functions of the time and the radial coordinate. They are defined as
\begin{equation}
    H^{\ell,m} := \int H(\theta,\phi)_{-2}\bar{Y}^{\ell,m}(\theta,\phi)\dd\Omega,
\end{equation}
where $_{-2}\bar{Y}^{\ell,m}(\theta,\phi)$ is the complex conjugate of $_{-2}Y^{\ell,m}(\theta,\phi)$ and $\dd\Omega$ represents the integral over the solid angle for $(\theta,\phi) \in [0,\pi]\times[0,2\pi)$.

The spin-weighted spherical harmonics can be explicitly expressed as~\cite{goldberg_macfarlane_1967}
\begin{align}\label{eq:sphham}
    \nn _sY^{\ell,m}(\theta,\phi) =& \sqrt{\frac{(\ell+m)!(\ell-m)!(2\ell+1)}{4\pi(\ell+s)!(\ell-s)!}}e^{im\phi} \\
    \nn &\sum_{k = 0}^{\ell-s}(-1)^{l-k-s+m}\left(\begin{array}{c} \ell-s \\ k \end{array}\right)\left(\begin{array}{c} \ell+s \\ k+s-m \end{array}\right) \\
    &\cos^{2k+s-m}\left(\frac{\theta}{2}\right)\sin^{2\ell-2k-s+m}\left(\frac{\theta}{2}\right),
\end{align}
where
\begin{equation}
    \left(\begin{array}{c} n \\ k \end{array}\right) = \begin{cases} \frac{n!}{(n-k)!k!}, & \text{if $0 \leq k < n$} \\ 0, & \text{otherwise} \end{cases}
\end{equation}
are the binomial coefficients and $n!$ is the factorial of $n$.

The spin-weighted spherical harmonics fulfill the two following differential properties~\cite{ruiz_alcubierre_2008}
\begin{align}\label{eq:shdifz}
    J_z\,_sY^{\ell,m} =& im_sY^{\ell,m}, \\ \label{eq:shdifo}
    J_\pm\,_sY^{\ell,m} =& i\sqrt{(\ell \mp m)(\ell + 1 \pm m)}_sY^{\ell,m\pm 1},
\end{align}
where $J_z := \partial_\phi$ and $J_\pm := e^{\pm i\phi}(\pm i\partial_\theta - \cot(\theta)\partial_\phi - is\csc(\theta))$, and the identity
\begin{equation}\label{eq:shunit}
    \int\,_sY^{\ell,m}(\theta,\phi)\,_{s'}\bar{Y}^{\ell',m'}(\theta,\phi)\dd\Omega = \delta_{s,s'}\delta_{\ell,\ell'}\delta_{m,m'},
\end{equation}
where $\delta_{a,b}$ is the Kronecker-delta. Further, we have for the complex conjugate of a spin-weighted spherical harmonic
\begin{equation}\label{eq:shconj}
    _s\bar{Y}^{\ell,m}(\theta,\phi) = (-1)^{s+m}\,_{-s}Y^{\ell,-m}(\theta,\phi).
\end{equation}

\section{The effect of aberration and polarization rotation on the wave}\label{sec:eap}

As shown in Ref.~\cite{torres-orjuela_chen_2019} GWs emitted by a source moving with a constant velocity are affected by aberration and polarization rotation. However, the picture as discussed in Ref.~\cite{torres-orjuela_chen_2019} is not complete. Only the effect of the motion on the the orientation of the wave vector and the polarization directions is considered but not the effect on the modes. The description of the modes is necessary to capture properties of the gradient of the metric, which corresponds to the gravitational field~\cite{misner_thorne_1973}.

Our goal is to describe the modes seen by a distant observer for given modes in the rest frame of the source and a known velocity of the source relative to the observer. For this purpose, we consider how the complex amplitude of a moving source transforms when seen by a distant observer and solve for the modes the resting observer detects. We will see that this transformation and recalculation of the modes induces a mixture of the modes which then in turn affects the evolution of the particular modes.

The transformation of the gravitational radiation at a fixed time can be described by the aforementioned aberration and polarization rotation, when applying them to each ray. In this section, we will derive mathematical expressions for these two effects and then show how they translate into effects on the GW complex amplitude. We use that when considering these effects ray by ray, each ray is only affected by these effects (up to now there is no mode mixture which affects the modes). However, this transformation leads to the complex amplitude being described relative to a non-rectilinear coordinate system (CO) which differs from the one an observer would use. Therefore, when expressing this complex amplitude relative to the CO of the observer we will get a `new' complex amplitude, with modes different from the original ones.

Before deriving how aberration and polarization rotation affect the complex amplitude, let us establish appropriate COs. The decomposition in spin-weighted spherical harmonics is usually performed around the source. Therefore, we set a CO attached to the source's CoM so that the $z'$-axis is perpendicular to the orbital plain and the $x'$- and $y'$-axes lie in the orbital plain. We denote the polar angle, measured relative to the $z'$-coordinate, by $\theta'$ and the azimuthal angle, measured from the $x'$-coordinate, by $\phi'$. Moreover, we consider an observer far enough from the source so that the space-time around him is flat except for the GWs and assume the CoM of the source to be moving with a velocity $\bs{v} = (v_x,v_y,v_z)$ relative to this observer. As shown in Ref.~\cite{boyle_2016}, a rotation of the coordinate system can affect the modes but in a different manner as a velocity. Therefore, for simplicity and without restriction of generality, we set the COs of the observers $(t,x,y,z)$ or $(t,r,\theta,\phi)$ to be parallel to the COs of the source in the limit of vanishing velocity.

Last, we would like to mention that in the case of a time-dependent motion, more effects than aberration and polarization rotation can appear (see, e.g., Ref.~\cite{torres-orjuela_chen_2020a} for a time-dependent phase shift induced by aberration) but, for simplicity, we do not consider them and focus on the effects of a constant velocity.

\subsection{Aberration}\label{ssuc:abe}

Like for light, the velocity of GWs is finite but equal for all observers. Therefore, for an observer the same GW ray points in different directions when the source is moving as when the source is at rest. This effect, known as `aberration' for light~\cite{jackson_2009}, changes the perceived shape of a GW source and thus its decomposition in modes.

In the observer and the source frames the direction vectors (the spatial parts of the wave vector) of a GW ray can be described, respectively, by the radial vectors $\bs{e}_r := (\sin(\theta)\cos(\phi),\sin(\theta)\sin(\phi),\cos(\theta))$ and $\bs{e}'_r := (\sin(\theta')\cos(\phi'),\sin(\theta')\sin(\phi'),\cos(\theta'))$, respectively. Using a Lorentz transformation and normalizing the resulting vector, we find that the two vectors are related by~\cite{torres-orjuela_chen_2020a}
\begin{equation}
    \bs{e}'_r = \frac{\bs{e}_r - \gamma\bs{v} + \gamma^2\langle\bs{e}_r,\bs{v}\rangle\bs{v}/(\gamma+1)}{\gamma(1 - \langle\bs{e}_r,\bs{v}\rangle)},
\end{equation}
where $\gamma := (1-v^2)^{-1/2}$ is the Lorentz factor and $\langle\cdot,\cdot\rangle$ is the three dimensional Euclidean scalar product.

The aberration of the ray makes that the direction vector pointing towards $(\theta,\phi)$ in the observer frame points towards $(\theta',\phi')$ in the source frame. The angles in the source frame can be computed using the direction vector, $\bs{e}'_r$, as
\begin{align}\label{eq:abethe}
    \cos(\theta') =& (\bs{e}'_r)_z = \mathcal{D}(\theta,\phi) (\cos(\theta) + \mathcal{C}(\theta,\phi)v_z), \\ \label{eq:abephi}
    \tan(\phi') =& \frac{(\bs{e}'_r)_y}{(\bs{e}'_r)_x} = \frac{\sin(\theta)\sin(\phi) + \mathcal{C}(\theta,\phi)v_y}{\sin(\theta)\cos(\phi) + \mathcal{C}(\theta,\phi)v_x},
\end{align}
where $\mathcal{C}(\theta,\phi) := \gamma^2\langle\bs{e}_r,\bs{v}\rangle/(\gamma+1) - \gamma$ and $\mathcal{D}(\theta,\phi) := 1/\gamma(1 - \langle\bs{e}_r,\bs{v}\rangle)$. Note that $\mathcal{C}$ and $\mathcal{D}$ are functions of the spherical coordinates because of the projection of the velocity vector, $\bs{v}$, on the radial vector, $\bs{e}_r$.

\subsection{Polarization rotation}\label{ssec:pro}

The geometry of GWs is described by the aforementioned direction vector and the directions of the polarizations. The latter are represented by two spatial vectors of unit length that are perpendicular to each other and the direction vector. Lorentz transformations do not preserve angles in space. Therefore, the directions of the polarizations do not transform according to a Lorentz transformation between a moving source and one at rest~\cite{thorne_1987}. Their proper transformation is shown in \eqs{eq:pptrafo}{eq:cptrafo} and in fact induces a rotation of the directions of the polarizations in the plane perpendicular to the direction vector. Such a rotation will also affect the spherical properties of the GWs.

In the observer frame the directions of the polarizations can be represented by the two spatial 4-vectors $\hat{e}_\theta := (0,\cos(\theta)\cos(\phi), \cos(\theta)\sin(\phi), -\sin(\theta))$ and $\hat{e}_\phi := (0,-\sin(\phi), \cos(\phi), 0)$. For a source moving with a 3-velocity $\bs{v}$, or accordingly a 4-velocity $\hat{u} = \gamma(1,\bs{v})$, the directions of the polarizations have the form~\cite{thorne_1987}
\begin{align}\label{eq:pptrafo}
    \hat{e}'_\theta =& \hat{e}_\theta - \frac{\eta(\hat{e}_\theta,\hat{u})}{\eta(\hat{k},\hat{u})}\hat{k}, \\ \label{eq:cptrafo}
    \hat{e}'_\phi =& \hat{e}_\phi - \frac{\eta(\hat{e}_\phi,\hat{u})}{\eta(\hat{k},\hat{u})}\hat{k}.
\end{align}
Here $\eta(\cdot,\cdot)$ is the Minkowsky product, $\hat{k} := \hat{e}_t + \hat{e}_r$ is the normalized wave vector and $\hat{e}_t := (1,0,0,0)$ is the time vector. Note that $\hat{e}'_\theta$ and $\hat{e}'_\phi$ do not differ from $\hat{e}_\theta$ and $\hat{e}_\phi$, respectively, by a Lorentz transformation but through a transformation imposed by their basic properties. The polarization rotation is then the angle by which $\hat{e}'_{\theta,\phi}$ is rotated relative to $\hat{e}''_{\theta,\phi} := \Lambda(\bs{v})\hat{e}_{\theta,\phi}$, i.e., the Lorentz transformed polarization directions, in the plane perpendicular to the GW direction vector. Note that $\hat{e}_{\theta,\phi}$ transform like contravariant vectors because they are the basis of the GW tensor.

The $\hat{e}'_{\theta,\phi}$ and $\hat{e}''_{\theta,\phi}$ are all perpendicular to the 4-velocity, $\hat{u}$, and hence spatial vectors in the source frame. Moreover, they are all vectors of unit length. Therefore, $\hat{e}'_\theta$ and $\hat{e}''_\theta$ as well as $\hat{e}'_\phi$ and $\hat{e}''_\phi$ can only differ from each other by spatial rotations
\begin{align}
    \hat{e}'_\theta = R_\phi(\beta_\theta)R_r(\alpha_\theta)\hat{e}''_\theta, \\
    \hat{e}'_\phi = R_\theta(\beta_\phi)R_r(\alpha_\phi)\hat{e}''_\phi,
\end{align}
where $R_i(\alpha)$ is a rotation by an angle $\alpha$ along the vector $\hat{e}''_i$ for $i = r,\theta,\phi$.

The $\beta_{\theta,\phi}$ are related to rotations due to the aberration. Therefore, we focus on the $\alpha_{\theta,\phi}$, which represent the rotations of the polarization in the plane perpendicular to the direction vector. Using that $\hat{e}'_{\theta,\phi}$ and $\hat{e}''_{\theta,\phi}$ are all spatial vectors of unit length we get that $\langle\bs{e}'_\theta,\bs{e}''_\phi\rangle = -\sin(\alpha_\theta)$ and $\langle\bs{e}'_\phi,\bs{e}''_\theta\rangle = -\sin(\alpha_\phi)$. Using that for purely spatial vectors $\langle\bs{e}'_\theta,\bs{e}''_\phi\rangle = \eta(\hat{e}'_\theta,\hat{e}''_\phi)$ and $\langle\bs{e}'_\phi,\bs{e}''_\theta\rangle = \eta(\hat{e}'_\phi,\hat{e}''_\theta)$, we find
\begin{align}\label{eq:polrot}
    \sin(\alpha_\theta) =& \frac{\gamma}{\gamma+1}\frac{\langle\bs{v},\bs{e}_\theta\rangle\langle\bs{v},\bs{e}_\phi\rangle}{1-\langle\bs{v},\bs{e}_r\rangle}, \\
    \sin(\alpha_\phi) =& \frac{\gamma}{\gamma+1}\frac{\langle\bs{v},\bs{e}_\theta\rangle\langle\bs{v},\bs{e}_\phi\rangle}{1-\langle\bs{v},\bs{e}_r\rangle}.
\end{align}
In particular, we find that both vectors are rotated by the same angle $\alpha := \alpha_\theta = \alpha_\phi$, which is what we expect for a rotation of the polarization that preserves its properties. Note that because of the projection of the velocity, $\bs{v}$, on the coordinate vectors, $\bs{e}_{r,\theta,\phi}$, the rotation angle is a function of the spherical coordinates, $\alpha = \alpha(\theta,\phi)$.

\subsection{The effect on the complex amplitude of the wave}\label{ssec:eow}

Aberration and polarization rotation affect the complex amplitude of the wave. Therefore, for a distant observer a GW source moving appears different from one at rest. In this section we relate the complex amplitude of a moving source seen by a distant observer, $H'(\theta,\phi)$, to the complex amplitude of the source in its own rest frame, $H(\theta',\phi')$.

The complex amplitude is a function of the spherical coordinates. Therefore, aberration causes that the observer sees the complex amplitude as
\begin{equation}\label{eq:comampab}
H'(\theta,\phi) = \mathcal{A}H(\theta',\phi'),
\end{equation}
where $\mathcal{A}$ denotes the transformation of the spherical coordinates induced by aberration and described in \eqs{eq:abethe}{eq:abephi}. We, further, note that the transformation $\mathcal{A}$ is itself a function of the spherical coordinates and hence cannot be described by a global transformation of the CO.

We know that GWs are not only affected by aberration but also by polarization rotation. A rotation of the basis vectors of the polarization in the same plane by an angle $\alpha$ (cf. \eq{eq:polrot}). This rotation means the polarizations combine to form `new' polarizations
\begin{align}
    \nn h'_+(\theta,\phi) =& \cos(2\alpha(\theta,\phi))\mathcal{A}h_+(\theta',\phi')\\
    &- \sin(2\alpha(\theta,\phi))\mathcal{A}h_\times(\theta',\phi'), \\
    \nn h'_\times(\theta,\phi) =& \sin(2\alpha(\theta,\phi))\mathcal{A}h_+(\theta',\phi') \\
    &+ \cos(2\alpha(\theta,\phi))\mathcal{A}h_\times(\theta',\phi').
\end{align}
Using Euler's equation, we then find for the complex amplitude
\begin{equation}\label{eq:comamptr}
    H'(\theta,\phi) = \mathcal{P}\mathcal{A}H(\theta',\phi'),
\end{equation}
where $\mathcal{P} := e^{-2i\alpha(\theta,\phi)}$. Note that because the polarization rotation is a function of the spherical coordinates, too, it cannot be described by a global transformation of the CO and, in particular, the phase shift does not represent a constant phase shift.

We find that the complex amplitude of the moving source transforms to the CO of the distant observer in a way that depends on the spherical coordinates $(\theta,\phi)$. Therefore, the decomposition in spherical modes of the moving source in the observer's rest frame differs from the decomposition in its own rest frame. We will elaborate this step in the next section.

\section{Excitation of the modes}\label{sec:eom}

\begin{figure*}[tpb] \centering \includegraphics[width=0.98\textwidth]{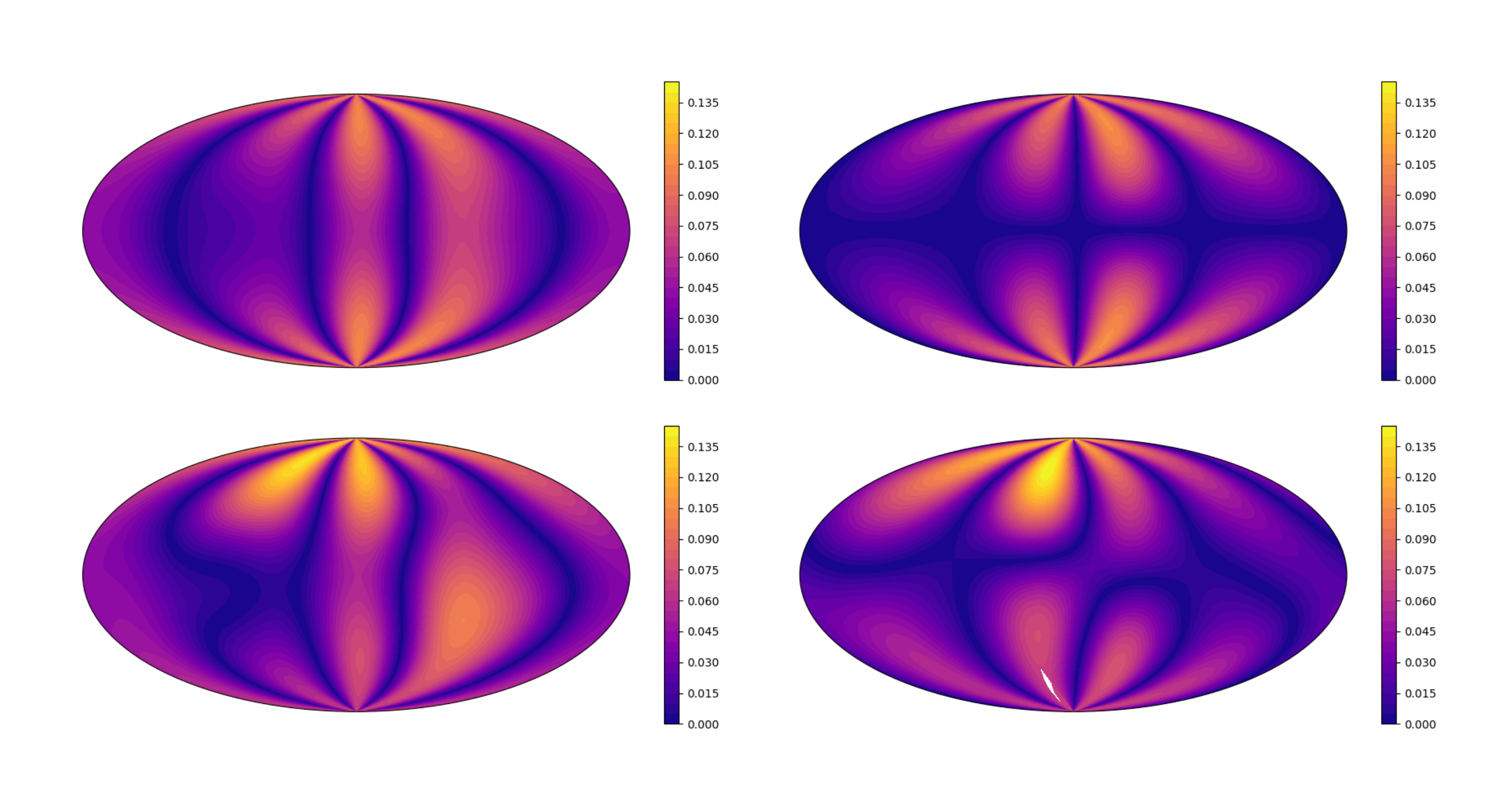}
\caption{
    Radiation patterns for the $+$-polarization (left) and the $\times$-polarization (right) of a BBH at rest (upper panels) and one moving with a constant velocity (lower panels) at the time of merger. In both cases the source is formed by two non-spinning BHs of the same mass on a non-eccentric orbit. The moving source has a velocity of 5\,\% the speed of light pointing along the diagonal of the $x$-$z$-plane.
}\label{fig:patt}
\end{figure*}

The modes of GWs represent the decomposition of the wave according to its spherical properties. A motion of the CoM of the source can affect the shape of the source through aberration and polarization rotation, as depicted in \fref{fig:patt}, thus changing these modes. In this section we show how the GW modes apparent to an observer change with the velocity of the source.

For the complex amplitude of the moving source as seen by the observer, we have
\begin{equation}\label{eq:dcamsof}
    H'(\theta,\phi) = \mathcal{T}H(\theta',\phi'),
\end{equation}
where $\mathcal{T} := \mathcal{P}\mathcal{A}$ means the transformations due to polarization rotation and aberration combined. Our goal is to derive explicit expressions for the modes of the moving source, $H'^{\ell,m}$, in terms of the modes of the source at rest, $H^{\ell,m}$, and the velocity of the source, $\bs{v}$, up to leading order in the amplitude of the velocity, $v$. Such an expansion is physically well motivated, since we expect most of the sources to have CoM velocities much smaller than the speed of light.

We start by expanding \eq{eq:dcamsof} using the coordinates $(\theta',\phi')$. We only will replace the primed coordinates by the observer's coordinates, $(\theta,\phi)$, after the expansion, to simplify the notation. However, this is just a matter of notation and does not place a restriction on the results. After expanding, we get
\begin{align}
    \nn H&'(\theta,\phi) \approx \left[\mathcal{T}H(\theta',\phi')\right]_{v=0} + \bigg[\mathcal{P}(\partial_{\theta'}\mathcal{A} H(\theta',\phi'))\frac{\dd\theta'}{\dd v} \\
    &+ \mathcal{P}(\partial_{\phi'}\mathcal{A} H(\theta',\phi'))\frac{\dd\phi'}{\dd v} + (\partial_{\alpha}\mathcal{P})\mathcal{A}H(\theta',\phi')\frac{\dd\alpha}{\dd v}\bigg]_{v=0}v.
\end{align}

Using that $\theta'|_{v=0} = \theta$, $\phi'|_{v=0} = \phi$ and $\alpha|_{v=0} = 0$, we get
\begin{align}
    \nn H'(\theta&,\phi) \approx H(\theta,\phi) + \bigg(\partial_{\theta} H(\theta,\phi)\frac{\dd\theta'}{\dd v}|_{v=0} \\
    &+ \partial_{\phi} H(\theta,\phi)\frac{\dd\phi'}{\dd v}|_{v=0} - 2iH(\theta,\phi)\frac{\dd\alpha}{\dd v}|_{v=0}\bigg)v.
\end{align}
From \eq{eq:polrot} we see that the rotation of the polarization only enters to the third order of $v$ and hence we can ignore the last term in the previous equation. We, further, replace $\theta'$ and $\phi'$ using \eqs{eq:abethe}{eq:abephi} and expand again to linear order in $v$, to find
\begin{align}\label{eq:decpri}
    \nn H'(\theta,\phi) \approx& H(\theta,\phi) + \frac{1}{\sin(\theta)}\big[(\partial_{\theta} H(\theta,\phi))(v_z - v_r\cos(\theta)) \\
    &+ (\partial_{\phi} H(\theta,\phi))(v_x\sin(\phi) - v_y\cos(\phi))\big],
\end{align}
where $v_r := \langle\bs{v},\bs{e}_r\rangle$.

Using the decomposition of $H$ in terms of spin-weighted spherical harmonics in \eq{eq:deccoma} and the differential properties in \eqs{eq:shdifz}{eq:shdifo}, we then find
\begin{equation}\label{eq:capredec}
    H'(\theta,\phi) \approx \sum_{\ell = 2}^\infty \sum_{m = -\ell}^\ell (1 + f_{\ell,m}(\theta,\phi))H^{\ell,m}\,_{-2}Y^{\ell,m}(\theta,\phi),
\end{equation}
where
\begin{align}\label{eq:fcoef}
    \nn f_{\ell,m}(\theta,\phi) :=& \frac{1}{\sin(\theta)}\bigg[im(v_z - v_r\cos(\theta)) \\
    &+ \frac12 A(\ell,m)(v_x\sin(\phi)-v_y\cos(\phi))\bigg]
\end{align}
and
\begin{align}
    \nn A(\ell,m) :=& \sqrt{(\ell-m)(\ell+1+m)} \\
    &- \sqrt{(\ell+m)(\ell+1-m)}.
\end{align}

The $f_{\ell,m}$ are functions of the spherical coordinates. Therefore, we need to further decompose \eq{eq:capredec} in order to get expressions in terms of only the spin-2 spherical harmonics and functions of only the time and radial coordinates. We decompose $H'$ in terms of spin-2 spherical harmonics, thus finding
\begin{equation}\label{eq:mogwde}
    H'(\theta,\phi) = \sum_{\ell = 2}^\infty \sum_{m = -\ell}^{\ell} H'^{\ell,m}\,_{-2}Y^{\ell,m}(\theta,\phi),
\end{equation}
where
\begin{equation}
    H'^{\ell,m} = \int H'(\theta,\phi)\,_{-2}\bar{Y}^{\ell,m}(\theta,\phi)\dd\Omega.
\end{equation}
Using \eqs{eq:shunit}{eq:shconj}, we then get
\begin{align}\label{eq:bemoin}
    \nn H&'^{\ell,m} = H^{\ell,m} + (-1)^{m}\sum_{\ell' = 2}^\infty\sum_{m' = -\ell'}^{\ell'} H^{\ell',m'} \\
    &\times\int f_{\ell',m'}(\theta,\phi)\,_{-2}Y^{\ell',m'}(\theta,\phi)\,_2Y^{\ell,-m}(\theta,\phi)\dd\Omega.
\end{align}

The integral in \eq{eq:bemoin} can be solved analytically using the expression in \eq{eq:sphham} and by decomposing it in its partial integrals, to find
\begin{align}\label{eq:decexi}
    \nn H'&^{\ell,m} = H^{\ell,m} - \frac{(-1)^{m}}{4^{\ell+1}i} \Bigg[\sum_{\ell_0 = \text{max}(2,|m|)}^\infty\frac{H^{\ell_0,m}}{4^{\ell_0}}C_0(\ell_0,\ell,m) \\
    \nn &+ \sum_{\ell_+ = \text{max}(2,|m+1|)}^\infty\frac{H^{\ell_+,m+1}}{4^{\ell_+}}C_+(\ell_+,\ell,m) \\
    &+ \sum_{\ell_- = \text{max}(2,|m-1|)}^\infty \frac{H^{\ell_-,m-1}}{4^{\ell_-}}C_-(\ell_-,\ell,m)\Bigg].
\end{align}
Explicit expressions for $C_0(\ell_0,\ell,m)$, $C_+(\ell_+,\ell,m)$ and $C_-(\ell_-,\ell,m)$ can be found in the appendix.

A remarkable feature in \eq{eq:decexi} is that a mode from a moving source of the order $(\ell,m)$ only has contributions from the modes of a source at rest of the orders $m-1$, $m$ and $m+1$ but of all orders $\ell' = \text{max}(2,|m-1|),...,\infty$. We expect contributions from modes of the orders $m \pm p$ when including terms of the order $v^p$. However, why the the order $m$ scales with the order of the velocity, while the order $\ell$ is independent of the order of the velocity, remains a question that requires further analysis.

We point out that the approach we use here implicitly assumes a decomposition of GWs in their modes at a fixed time, since we consider the effects of velocity on the direction vector and the polarization vectors but not on the frequency. The frequency of GWs is affected by a velocity through Doppler shift, which is also direction-dependent and hence could affect the decomposition of GWs in modes. However, to consider the effect on the frequency would require to account for an evolution in time, which lies beyond the scope of this work. To detect the change in the evolution, an observation period of the order of $\omega/\Delta\omega$, where $\omega$ is the frequency of the GW and $\Delta\omega$ is the maximal difference in the frequency induced by the Doppler shift, is needed, which for typical velocities ($\lesssim1\,\%$ the speed of light) is of the order of several hundred cycles and hence much longer than the typical duration of a LIGO/Virgo event~\cite{GWTC1,GWTC2}. Therefore, our approach represents an important step in understanding the effect of velocity on the modes of GWs but more work considering the Doppler shift needs to be conducted in order to obtain a more complete picture. Despite the restrictions of our approach, we will analyse in the remainder of this paper the consequences of our results for GWs taking into account their evolution in time. Although, the results obtained may not represent the complete picture they are still valid and can be considered as a minimal case. In particular, we expect that by considering the Doppler effect the modes from a moving source and one at rest will differ even more.

\begin{figure}[tpb] \centering \includegraphics[width=0.48\textwidth]{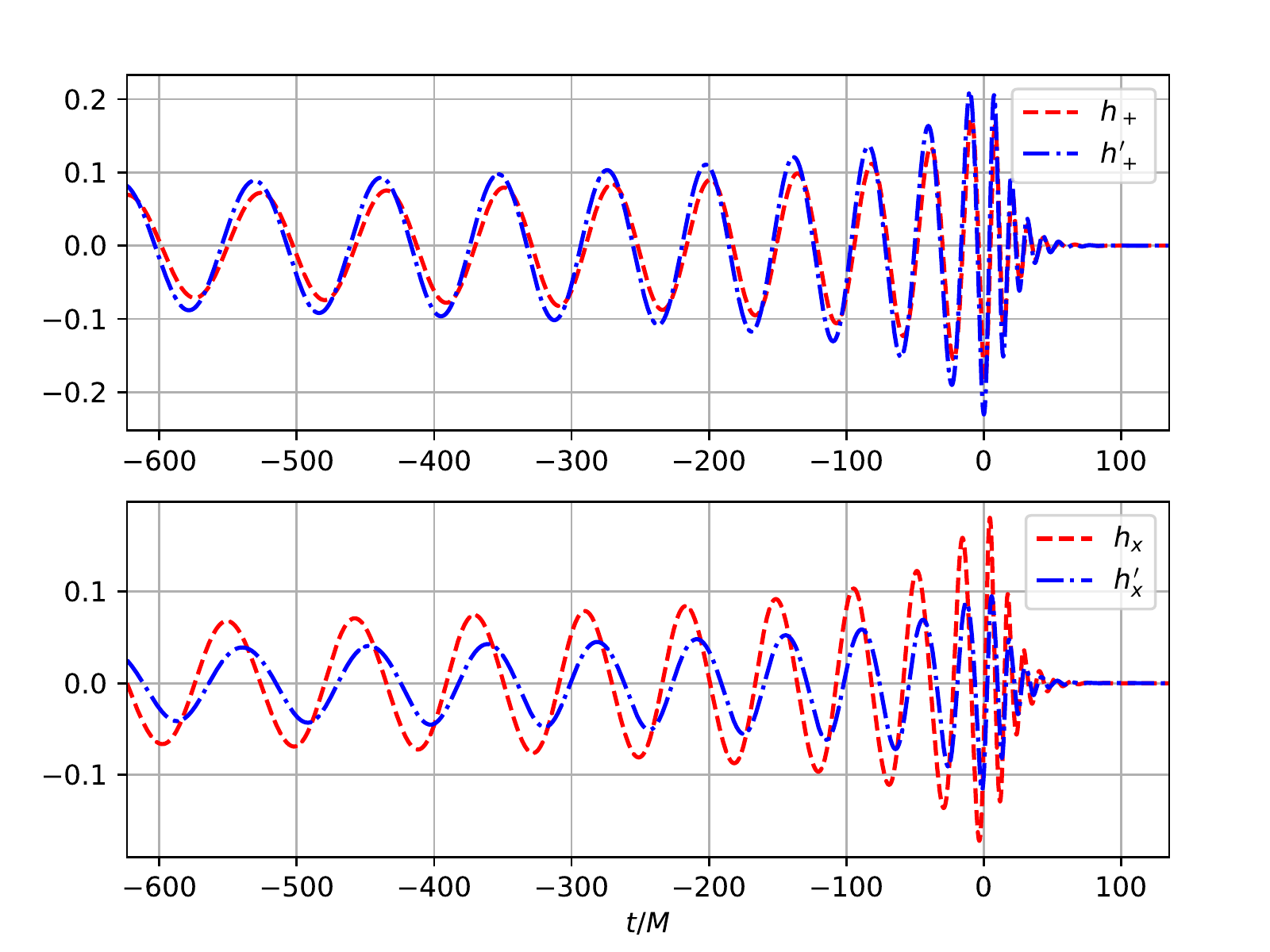}
\caption{
    The $+$- and $\times$-polarizations of a moving source, $h'_{+,\times}$, and one at rest, $h_{+,\times}$. For both cases we consider equal mass binaries with a total mass $M$ in the observer frame of non-spinning black holes on non-eccentric orbits. Both sources are seen by an observer at the sky location $(\theta,\phi) = (45^\circ,0^\circ)$ in the source frame. The velocity points in the direction $(\theta_v,\phi_v) \approx (55^\circ,45^\circ)$ and has a magnitude of $v = 0.1\,c$ to make the effects visible by eye.}
\label{fig:wafo}
\end{figure}

In \fref{fig:wafo} we compare the polarizations of a moving source, $h'_{+,\times}$, and those of a source at rest, $h_{+,\times}$. In the upper plot we see that the amplitude of the $+$-polarization for the moving source is enhanced relative to the amplitude of the source at rest. However, in the lower plot we see that for the $\times$-polarization the amplitude of the moving source is suppressed relative to the amplitude of the non-moving source. Note, that this is not a general feature but depends on the specific properties of the source, the velocity and the viewing angle. Moreover, the frequency of the GWs from the
moving source is shifted by a time-dependent factor relative to the frequency of the GWs from the source at rest. This shift appears because the overall frequency of GWs depends on the relative contribution of the particular modes, which for a moving source is different from one at rest. We discuss this effect in more detail in \sref{sec:fresh}.

\section{Breaking the degeneracy between Doppler shift and mass}\label{sec:bddsm}

It is well known that, when only considering the effect of motion on the frequency of GWs, a moving source is degenerate with a source with identical parameters but with a different mass~\cite{schutz_1986,cutler_flanagan_1994}. This circumstance is usually denoted as the mass-redshift degeneracy of GWs. However, in the previous section we showed that when considering aberration and polarization rotation, not only the frequency of the GWs changes but also the amplitude of their spherical modes. This change of the modes leads to a change in the amplitude of the polarizations and a time-dependent frequency shift, which makes it possible to differ between a moving source and one at rest.

One now might think that a moving source and one at rest remain degenerate if one not only adjust their mass to account for the Doppler shift but also their orientation to account for the effects of aberration and polarization rotation. In this section we show this is \textit{not true}. That means we show that the GWs from a moving source differ from those from a source at rest, regardless of any variation of the mass and orientation.

To show the difference between the resting and the moving source, we consider two sources with identical properties but the source at rest having a mass $M$ and the moving source a mass $M/\mathcal{D}(\theta,\phi)$. Here, $\mathcal{D}(\theta,\phi)$ is the Doppler shift in direction $(\theta,\phi)$ and its correction ensures that the time scales of both systems are the same in the frame of a distant observer. If the source at rest and the moving one indeed only would differ by an additional correction of their orientation, we would have
\begin{equation}\label{eq:assro}
    H(\tilde{\theta},\tilde{\phi}) = H'(\theta,\phi),
\end{equation}
where $(\tilde{\theta},\tilde{\phi})$ are two smooth functions of $(\theta,\phi)$ and the velocity of the source. By showing that this equation is \textit{not true}, we will prove that the two sources have different signals.

We start using the mode decomposition of the two complex amplitudes in \eq{eq:assro}, multiplying the equation by $_{-2}\bar{Y}^{k,n}(\theta,\phi)$ and integrating over $(\theta,\phi)$, to get
\begin{align}
    \nn &\sum_{\ell = 2}^\infty \sum_{m = -\ell}^{\ell} H^{\ell,m}\int\,_{-2}Y^{\ell,m}(\tilde{\theta},\tilde{\phi})_{-2}\bar{Y}^{k,n}(\theta,\phi)\dd\Omega = \\
    &\sum_{\ell = 2}^\infty \sum_{m = -\ell}^{\ell} H'^{\ell,m}\int\,_{-2}Y^{\ell,m}(\theta,\phi)_{-2}\bar{Y}^{k,n}(\theta,\phi)\dd\Omega.
\end{align}
Because $(\tilde{\theta},\tilde{\phi})$ are smooth functions of $(\theta,\phi)$, $_{-2}Y^{\ell,m}(\tilde{\theta},\tilde{\phi})$ can be decomposed in terms of spin-2 spherical harmonics over the basis $(\theta,\phi)$
\begin{equation}
    _{-2}Y^{\ell,m}(\tilde{\theta},\tilde{\phi}) = \sum_{a = 2}^\infty \sum_{b = -a}^{a} \mathcal{Y}^{\ell,m}_{a,b}\,_{-2}Y^{a,b}(\theta,\phi),
\end{equation}
where
\begin{equation}
    \mathcal{Y}^{\ell,m}_{a,b} := \int\,_{-2}Y^{\ell,m}(\tilde{\theta},\tilde{\phi})_{-2}\bar{Y}^{a,b}(\theta,\phi)\dd\Omega.
\end{equation}
Using this decomposition and \eq{eq:shunit}, we find
\begin{equation}
    \sum_{\ell = 2}^\infty \sum_{m = -\ell}^{\ell} H^{\ell,m}\mathcal{Y}^{\ell,m}_{k,n} = H'^{k,n}.
\end{equation}

Last, we replace $H'^{k,n}$ using \eq{eq:decexi} and get
\begin{align}\label{eq:proro}
    \nn \sum_{\ell = 2}^\infty &\sum_{m = -\ell}^{\ell} H^{\ell,m}\mathcal{Y}^{\ell,m}_{k,n} = H^{k,n} \\
    \nn &- \frac{(-1)^{n}}{4^{k+1}i}\Bigg[\sum_{\ell_0 = \text{max}(2,|n|)}^\infty\frac{H^{\ell_0,n}}{4^{\ell_0}}C_0(\ell_0,k,n) \\
    \nn &+ \sum_{\ell_+ = \text{max}(2,|n+1|)}^\infty\frac{H^{\ell_+,n+1}}{4^{\ell_+}}C_+(\ell_+,k,n) \\
    &+ \sum_{\ell_- = \text{max}(2,|n-1|)}^\infty \frac{H^{\ell_-,n-1}}{4^{\ell_-}}C_-(\ell_-,k,n)\Bigg].
\end{align}
We see that on the left hand side of this equation the index $n$ is independent of the modes of the source, $H^{\ell,m}$. In contrast on the right hand side the index $n$ depends on the modes of the source for all four terms. Therefore, for general sources and velocities \eq{eq:proro} is not fulfilled.

To make this more clear, we consider a hypothetical source with only the $(2,2)$ mode. In this case \eq{eq:proro} reduces to the form
\begin{align}\label{eq:proros}
    \nn \mathcal{Y}^{2,2}_{k,n} =& 1 + \frac{i}{1024}[C_0(2,k,2) - C_+(2,k,1) \\
    &- C_-(2,k,3)].
\end{align}
Here, the left hand side depends on $n$ through the $\mathcal{Y}^{2,2}_{k,n}$ while the right hand side, in contrast, is independent of $n$. Therefore, for \eq{eq:proros} to be fulfilled it would be necessary that the $\mathcal{Y}^{2,2}_{k,n}$ are independent of $n$. For that to happen would require that the velocity has no components perpendicular to the angular momentum of the source so that it is independent of $\phi$. However, for general sources and orientations of the velocity this is not true, thus implying that \eq{eq:proros} is in general not fulfilled. We note that the velocity being parallel to the angular momentum of the source is necessary but not sufficient to ensure that \eq{eq:proros} is fulfilled. Even in the case where the velocity has no components perpendicular to the angular momentum of the source the $\mathcal{Y}^{2,2}_{k,n}$ can depend on $n$.

We have shown that the GWs from two sources with the same properties, except that one is moving and the other at rest, and a correction of their masses to account for the Doppler shift, look different. This remains true even when assuming the observer can `adjust' the orientation of the two sources. Thereby we have proven that for moving sources the degeneracy between mass and Doppler shift can be broken when considering the higher modes of the source.

Note, that for our proof we assumed to have two sources with identical properties except for their velocity and mass. To detect the velocity with certainty would, however, require to extract the effect of the velocity independent of all other parameters. That means to be able to tell apart if a source is moving independent of the combination of other parameters like spin, mass ratio, eccentricity, etc. As with all parameters with no fundamental degeneracy we expect this to be possible~\cite{cutler_flanagan_1994}, though, it needs to be studied case by case for each possible source and parameter. A detailed study of this problem would go beyond the scope of this work. However, we have shown that a constant velocity can be detected for extreme mass-ratio inspirals~\cite{torres-orjuela_amaro-seoane_2020} and are looking into the problem for LIGO/Virgo sources, of which the results will be published elsewhere.

\section{Induced frequency shift}\label{sec:fresh}

GW detection is most sensitive to the phases of the polarizations, $\Phi_{+,\times}$~\cite{lindblom_owen_2008}. Therefore, any effect on the phases, or accordingly the overall frequencies, $\omega_{+,\times} := \dd\Phi_{+,\times}/\dd t$, is of particular interest. The frequencies of particular modes of GWs can be represented as combinations of integer multiples of a fundamental frequency, $\omega_o$~\cite{arun_buonanno_2009}. However, in the case of the overall frequencies $\omega_{+,\times}$ such a simple representation is not possible, because it depends on the contribution of the particular modes, i.e., their amplitude, to each polarization. This fact, on the other hand implies, that a change of the amplitude of the particular modes can lead to a change of the overall frequencies.

In this section we show how the overall frequencies in the observer frame, $\omega'_{+,\times}$, appear shifted relative to the overall frequencies in the source frame, $\omega_{+,\times}$, when the source is moving with a constant velocity, $\bs{v}$. We assume that for both polarizations the amplitude, $A_{+,\times}(t)$, and the overall frequency, $\omega_{+,\times}$, only change slowly with time, i.e., on time scales much bigger than $1/\omega_o$, and hence we can ignore their time derivatives. We further assume that the magnitude of the velocity is small, $v \ll 1$.

Before deriving the frequency shift, we would like to point out that for a moving source the fundamental frequency, $\omega_o$, will appear Doppler shifted to the observer, i.e., $\omega'_o = \mathcal{D}\omega_o$. However, a source which is at rest but $\mathcal{D}$ times more massive than the moving source emits GWs with the same fundamental frequency~\cite{chen_li_2017}. Therefore, we can treat the moving source and the source at rest as having the same fundamental frequency in the observer frame by assuming they have identical intrinsic properties but different masses.

The frequency of a wave is the inverse of the time required for one cycle, e.g., the time between two adjacent maxima of the wave. For $t^{(1)}_{+,\times}$ and $t^{(2)}_{+,\times}$ two adjacent maxima of $h_{+,\times}(t)$, we find for its frequency
\begin{equation}\label{eq:norfre}
    \omega_{+,\times} = \frac{2\pi}{t^{(2)}_{+,\times}-t^{(1)}_{+,\times}},
\end{equation}
where we assume that $t^{(2)}_{+,\times} > t^{(1)}_{+,\times}$. For a moving source we have different polarizations $h'_{+,\times}(t)$ with (slightly) different maxima $t'^{(1)}_{+,\times}$ and $t'^{(2)}_{+,\times}$ and hence we find for its frequency
\begin{equation}\label{eq:shifre}
    \omega'_{+,\times} = \frac{2\pi}{t'^{(2)}_{+,\times}-t'^{(1)}_{+,\times}}.
\end{equation}

According to \eqs{eq:mogwde}{eq:decexi} the polarizations of GWs from a moving source can be decomposed as
\begin{equation}\label{eq:hprispl}
    h'_{+,\times}(t,\theta,\phi) = h_{+,\times}(t,\theta,\phi) + \Delta h_{+,\times}(t,\theta,\phi),
\end{equation}
where $\Delta h_{+,\times}/h_{+,\times} \sim v$. Therefore, using that the magnitude of the velocity is small, we can write
\begin{equation}\label{eq:relmax}
    t'^{(1,2)}_{+,\times} = t^{(1,2)}_{+,\times} + \Delta t^{(1,2)}_{+,\times},
\end{equation}
where $\omega_{+,\times}\Delta t^{(1,2)}_{+,\times} \sim v$.

We use that the time derivatives of $h_{+,\times}(t)$ and $h'_{+,\times}(t)$ vanish at $t^{(1,2)}_{+,\times}$ and $t'^{(1,2)}_{+,\times}$, respectively. Then expanding to linear order in $\omega_{+,\times}\Delta t^{(1,2)}_{+,\times}$, we find 
\begin{equation}\label{eq:timesh}
    \Delta t^{(1,2)}_{+,\times} = -\frac{(\dd\Delta h_{+,\times}(t)/\dd t)|_{t=t^{(1,2)}_{+,\times}}}{(\dd^2 h_{+,\times}(t)/\dd t^2)|_{t=t^{(1,2)}_{+,\times}}},
\end{equation}
where we ignored $(\dd^2\Delta h_{+,\times}/\dd t^2)\Delta t^{(1,2)}_{+,\times}$ because it is of the order $v^2$.

Using that $h_{+,\times}(t)$ has maxima at $t^{(1,2)}_{+,\times}$, that its amplitude only changes slowly in time and \eq{eq:hprispl}, we get
\begin{align}
    \frac{\dd\Delta h_{+,\times}(t)}{\dd t}\bigg|_{t=t^{(1,2)}_{+,\times}} =& -A'_{+,\times}(t^{(1,2)}_{+,\times})\omega'_{+,\times}s^{1,2}_{+,\times}, \\
    \frac{\dd^2 h_{+,\times}(t)}{\dd t^2}\bigg|_{t=t^{(1,2)}_{+,\times}} =& -A_{+,\times}(t^{(1,2)}_{+,\times})\omega^2_+,
\end{align}
where $s^{(1,2)}_{+,\times} := \sin(\Delta\omega_{+,\times} t^{(1,2)}_{+,\times})$, and $\Delta\omega_{+,\times} := \omega'_{+,\times} - \omega_{+,\times}$ is the difference between the frequencies in the observer and source frame. Next, we use that because the amplitudes change slowly with time $A_{+,\times}(t^{(1)}_{+,\times}) \approx A_{+,\times}(t^{(2)}_{+,\times}) \approx A_{+,\times}(t)$ for $t^{(1)}_{+,\times} \lesssim t \lesssim t^{(2)}_{+,\times}$ and the same for $A'_{+,\times}(t)$, and expand $s^{(1,2)}_{+,\times}$ to linear order in $\Delta\omega_{+,\times}t^{(1,2)}_{+,\times}$, to find
\begin{equation}\label{eq:deltres}
    \Delta t^{(1,2)}_{+,\times} = -\frac{A'_{+,\times}(t)}{A_{+,\times}(t)}\frac{\omega'_{+,\times}}{\omega^2_{+,\times}}\Delta\omega_{+,\times}t^{(1,2)}_{+,\times}.
\end{equation}

Using \eqs{eq:norfre}{eq:relmax} together with \eq{eq:deltres} in \eq{eq:shifre}, expanding to linear order in $\Delta\omega_{+,\times}/\omega_{+,\times} \sim v$ and replacing back $\Delta\omega_{+,\times} = \omega'_{+,\times}-\omega_{+,\times}$, we find
\begin{equation}\label{eq:shifreres}
    \omega'_{+,\times} = \frac{A_{+,\times}(t)}{A'_{+,\times}(t)}\omega_{+,\times}.
\end{equation}
We point out, that we also get $\omega'_{+,\times} = \omega_{+,\times}$ as a possible solution. However, by decomposing the wave in its modes and changing their respective amplitude, we can confirm that \eq{eq:shifreres} is the general solution and $\omega'_{+,\times} = \omega_{+,\times}$ only is valid in some special cases.

Using \eq{eq:shifreres} together with \eqs{eq:decexi}{eq:hprispl}, we finally get
\begin{equation}\label{eq:freshire}
    \omega'_{+,\times} = (1 \mp v\alpha_{+,\times}(t))\omega_{+,\times},
\end{equation}
where 
\begin{align}\label{eq:alppl}
    \alpha_+(t) =& \frac{[\sum_{\ell = 2}^\infty\sum_{m = -\ell}^\ell\Delta H^{\ell,m}(t)\,_{-2}Y^{\ell,m}(\theta,\phi)]_\times}{A_+(t)}, \\ \label{eq:alpcr}
    \alpha_\times(t) =& \frac{[\sum_{\ell = 2}^\infty\sum_{m = -\ell}^\ell\Delta H^{\ell,m}(t)\,_{-2}Y^{\ell,m}(\theta,\phi)]_+}{A_\times(t)},
\end{align}
$\Delta H^{\ell,m}(t) := -i(H'^{\ell,m}(t) - H^{\ell,m}(t))/v$, $[f]_+$ means the amplitude of the real part of $f$ and $[f]_\times$ means minus the amplitude of the imaginary part of $f$.

In \eq{eq:freshire} we see that both frequencies are shifted proportional to the magnitude of the velocity, but with opposite sign. This shift in opposite directions can also
be seen in \fref{fig:wafo}, which was generated by only using the change of the amplitude of the particular modes and does not rely on \eq{eq:freshire}. Moreover, we see in \eqs{eq:alppl}{eq:alpcr} that the shift of the $+$-polarization depends on the change of the $\times$-polarization and vice versa. This is because the shift of the particular polarization is induced by a mixing with the other polarization. Last, we want to highlight that because the amplitude of the modes is time-dependent these frequency shifts are also time-dependent, even for constant velocities.

\section{Detectability}\label{sec:det}

We have shown that GWs emitted by a moving source differ from those emitted by a source at rest. Therefore, an observer using a waveform model not containing any information about the CoM velocity detects a difference to the incoming GWs from the moving source. This difference expresses as a reduced match, $M(h,h')$, or accordingly an increased mismatch $1-M(h,h')$, between the model waveform, $h$, and the incoming wave, $h'$. However, if the mismatch can be resolved depends on the loudness of the source, which is quantified by its signal-to-noise ratio (SNR), $\rho(h')$. In general a mismatch between two waveforms can be detected if the SNR fulfils~\cite{lindblom_owen_2008}
\begin{equation}
    \rho(h') > \frac{1}{\sqrt{2(1-M(h,h'))}}.
\end{equation}
Note that to be able to detect the mismatch of the two waveforms is only a necessary condition to detect the velocity. As discussed in \sref{sec:bddsm}, it is also necessary to exclude possible `confusions' between different parameters. Such an analysis goes beyond the scope of this work but is been studied by some of the authors and the results will be published elsewhere.

\begin{figure}[tpb] \centering \includegraphics[width=0.48\textwidth]{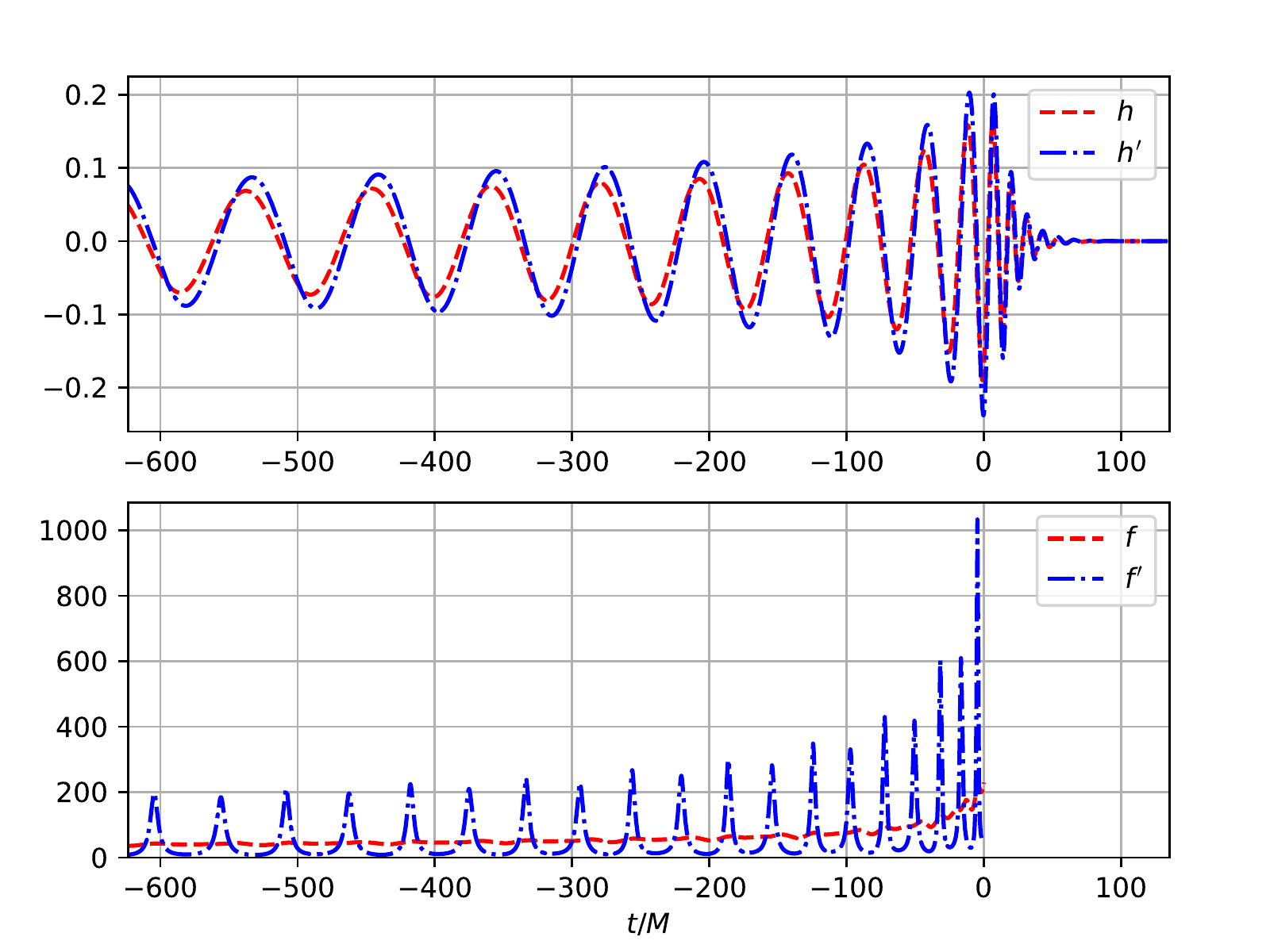}
\caption{
    The waveforms from \fref{fig:wafo} projected onto a LIGO like detector. The upper plot shows the signals of the source at rest, $h$, and the moving one, $h'$ detected by such a the detector for a configuration where the detector is equally sensitive to both polarizations. In the lower plot we see the corresponding frequencies of the source at rest, $f$, and of the moving source, $f'$, until merger.
    }
\label{fig:sig}
\end{figure}

Before analysing the detectability of the velocity in detail, let us have a look on the signals produced by the waveforms in \fref{fig:wafo}. In the upper part of \fref{fig:sig} we show the signals of a moving source, $h'$, and a source at rest, $h$, a LIGO like detector (without noise) would detect for a configuration where the detector is equally sensitive to both polarizations. For the moving source and the one at rest the dominant $+$-polarization differs less than the $\times$-polarization (cf. \fref{fig:wafo}) and hence the signals detected do not seem to differ a lot. However, when considering the frequencies of the signals, which we show in the lower part of \fref{fig:sig}, we see a remarkable difference. While the frequency of the source at rest, $f$, increases smoothly in time, the frequency of the moving source, $f'$, oscillates considerable. Such an oscillation is very different from what we would expect for an equal mass binary of non-spinning BHs on non-eccentric orbits, as considered in this case. The frequencies were computed using the function `\texttt{freqeuncy\_from\_polarizations}' from the PyCBC software package~\cite{pycbc_2020,pycbc_inference_2019} and are shown until merger time because after merger erroneous oscillations arise due to numerical inaccuracies.

We estimate what SNR is required to resolve the difference between the model waveform and the incoming wave. For this purpose we generate a model waveform, $h$, using the Numerical Relativity surrogate model `NRHybSur3dq8', which can generate waveforms containing the most important modes up to the $(5,5)$-mode for BBH of mass ratios, $q=m_1/m_2$, up to 8 ($m_1$ ($m_2$) being the mass of the heavier (lighter) BH)~\cite{varma_field_2019}. We generate the waveform of the incoming wave, $h'$, using again the modes obtained from NRHybSur3dq8 but distorting them according to \eq{eq:decexi}, where we set the velocity of the source to lie in the orbital plain of the source. Both sources are set to have a total mass of $40\,M_\odot$ in the observer frame so that they have the same fundamental frequency in this frame. Further, we consider a LIGO/Virgo like detector with a GW coming in perpendicular to its plane and the polarizations rotated by $22.5^\circ$ relative to its arms, in order for the detector to be equally sensitive to both polarizations~\cite{sathyaprakash_schutz_2009}. The waveforms are set to have an initial frequency of 50\,Hz and we compute their match, $M(h,h')$, using the `\texttt{match}' function from the PyCBC software package~\cite{pycbc_2020,pycbc_inference_2019}.

Note that the two sources only differ by their velocity and their mass in the source frame (as to have the same mass in the observer frame). All other intrinsic and extrinsic parameters are fixed to be equal for both sources. As discussed in \sref{sec:bddsm} the two signals still differ when allowing for a change in the orientation of the sources and, in general, we expect the sources to have different signals after changing different parameters since there is no fundamental degeneracy. However, in realistic (noisy) detections confusions could arise when changing several parameters at the same time. A study of this problem goes beyond the scope of this work but will be published elsewhere.

\begin{figure}[tpb] \centering \includegraphics[width=0.48\textwidth]{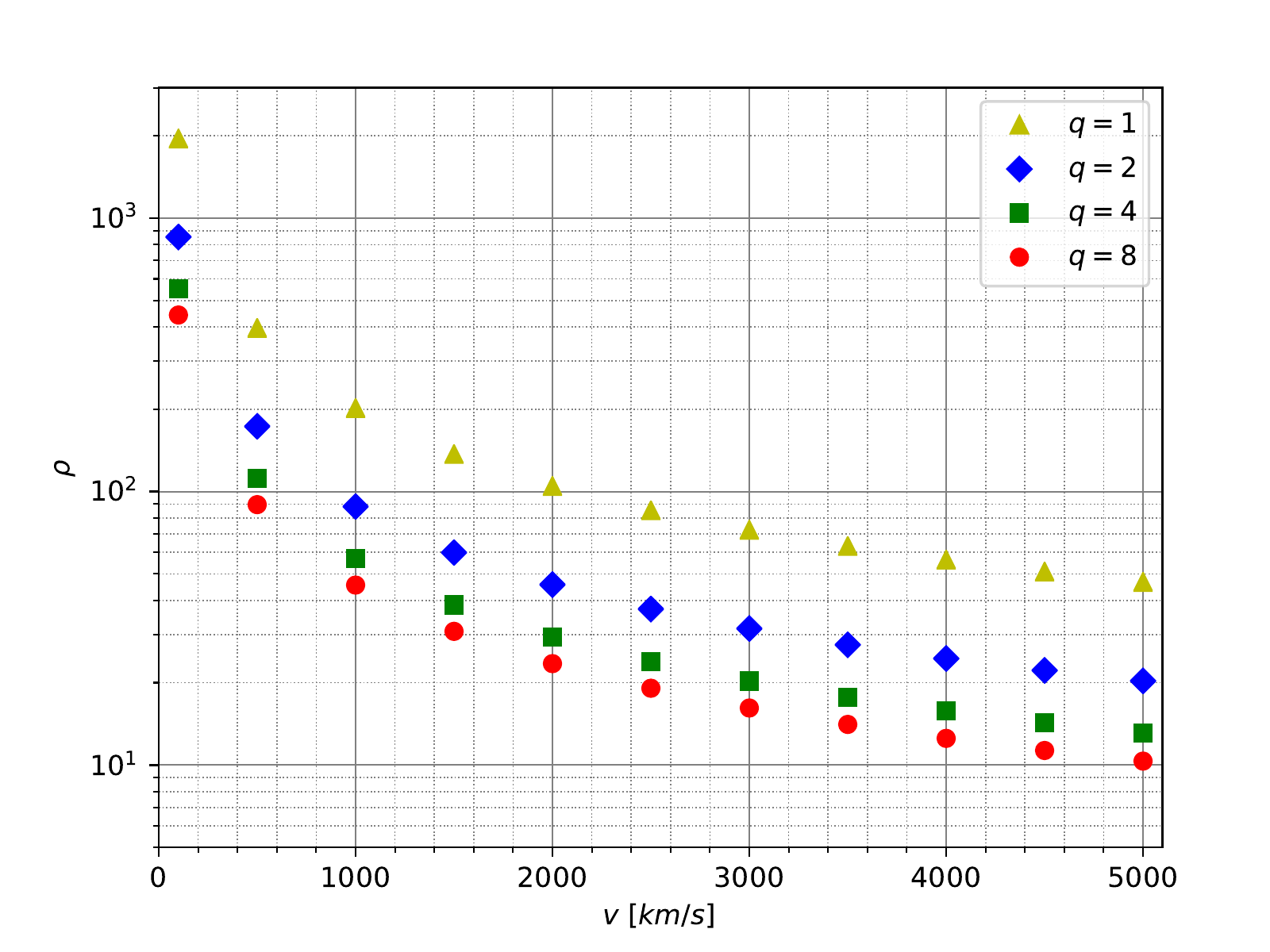}
\caption{
    SNR required to detect the mismatch between GWs from a moving source and one at rest, as a function of the velocity of the moving source. Different markers represent different mass ratios, $q$, for the merging BBH. For all cases the source is seen edge on ($\iota = 90^\circ$).
}\label{fig:snrm}
\end{figure}

\fref{fig:snrm} shows the SNR required to resolve the difference between the model and the incoming waveform for different mass ratios, $q=m_1/m_2$, as a function of the velocity of the source. For all cases we consider the source as being seen from edge on. We see that for increasing mass ratios a lower SNR is required to resolve the difference between the waveforms. This is because the velocity induces a frequency shift proportional to the subdominant modes of the GW, which are more prominent for sources of high mass ratios. In \fref{fig:snrm} we further see that for an SNR of around 20 and high mass ratios (cf. Ref.~\cite{ligo_2020a,ligo_2020b}) a constant velocity of only $2500\,\textrm{km\,s}^{-1}$ could be detected by LIGO/Virgo.

\begin{figure}[tpb] \centering \includegraphics[width=0.48\textwidth]{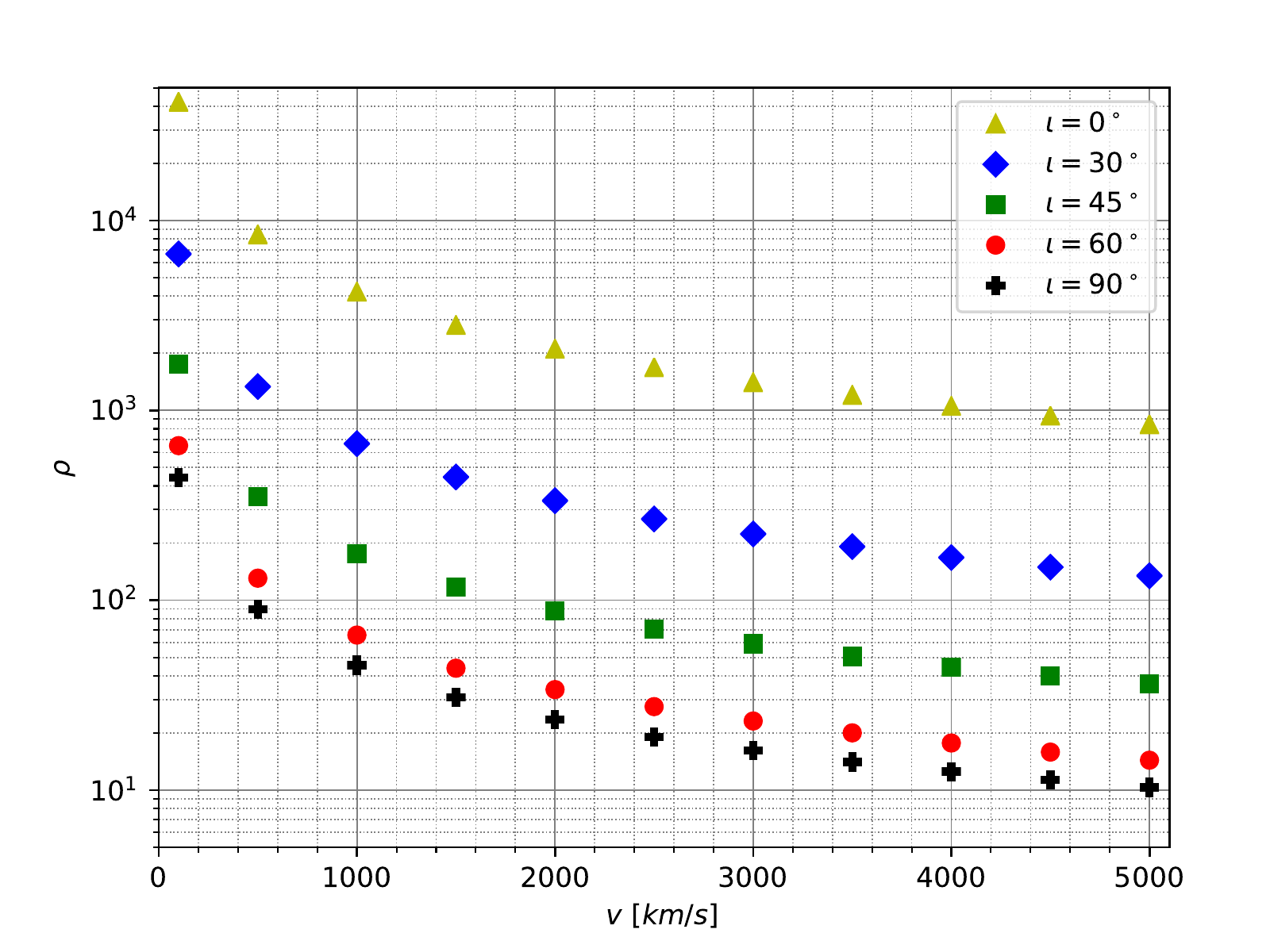}
\caption{
    SNR required to detect the mismatch between GWs from a moving source and one at rest, as a function of the velocity of the moving source. Different markers represent different inclinations of the source, $\iota$, relative to the observer, where $\iota = 0^\circ$ means face on. For all cases the merging BBH has a mass ratio $q = 8$.
}\label{fig:snri}
\end{figure}

In \fref{fig:snri} we show the SNR needed to detect the source's motion as a function of the magnitude of the velocity, for an observer seeing a source of mass ratio $q=8$ from different inclinations, $\iota$. The SNR required to resolve the motion is the lowest for a source being seen edge on ($\iota = 90^\circ$), where the subdominant modes are the strongest, and the highest when seen face on ($\iota = 0^\circ$), where the subdominant modes are the weakest~\cite{kalaghatgi_hannam_2020}. However, for inclinations higher then $45^\circ$ the SNR required to resolve the motion only differs by a small factor from the one required for a source being seen edge on. Only when the inclination goes below $45^\circ$ a significantly higher SNR is needed. 

In this section, we showed that considering the effect of a CoM velocity on the modes of GWs, constant velocities can be detected. The difference between a model waveform not including the effect of motion and an incoming wave from a moving source could be resolved by LIGO and Virgo for velocities of $2500\,\textrm{km\,s}^{-1}$, which is below the peculiar velocity of the fastest moving galaxies~\cite{scrimgeour_davis_2016}. Moreover, we showed that for sources with significant contributions from the subdominant modes (high mass ratios and inclinations) the effect of the velocity is more prominent.

\section{Conclusions}\label{sec:con}

We found that for a source of GWs moving with a constant velocity the amplitude of the modes change due to a mixing with other modes. We described the effect of a constant velocity on the complex polarization of GWs without restrictions on the velocity or the source and derived an analytic expression for the change of the modes to first order in the magnitude of the velocity. Moreover, we proved that considering the excitation of the modes the mass-redshift degeneracy of GWs is broken. We, further, showed that the excitement of the modes leads to a time-dependent frequency shift, where the shift for the frequency of the $+$-polarization depends on the change of the amplitude of the $\times$-polarization and vice versa.

We investigated the detectability of the induced frequency shifts and change of amplitudes. For this purpose we computed what SNR would be required to resolve the mismatch between a model waveform not including the effect of the motion and an incoming wave from a moving source, where both sources are scaled to have the same mass in the observer frame. We found that for an SNR of about 20, LIGO and Virgo could resolve constant velocities as low as $2500\,\textrm{km\,s}^{-1}$ when the source has a high mass ratio ($\approx 8$) and a high inclination ($\gtrsim 45^\circ$). In particular, we want to highlight that this effect breaks the degeneracy between mass and Doppler shift for GWs and represents the first method to detect the constant velocity of a source by only using GWs.

\section*{Acknowledgments}

This work is supported by the National Science Foundation of China grants No~11721303, 11873022, and 11991053. A.T.O. is partly supported by the Strategic Priority Research Program of the Chinese Academy of Sciences, Grant No.  XDB23040100 and No. XDB23010200. P.A.S. acknowledges support from the Ram{\'o}n y Cajal Programme of the Ministry of Economy, Industry and Competitiveness of Spain, as well as the COST Action GWverse CA16104.

\section*{Appendix}
\appendix

Here we give explicit expressions for the coefficients $C_0(\ell_0,\ell,m)$, $C_+(\ell_+,\ell,m)$ and $C_-(\ell_-,\ell,m)$ in \eq{eq:decexi}:
\begin{widetext}
\begin{align}\label{eq:coe0}
    \nn C_0(\ell_0,\ell,m) :=& 2\pi mv_z F(\ell_0,m;\ell,m)\sum_{k = \text{max}(0,m+2)}^{\ell_0+2}\sum_{a = 0}^{2\ell_0-2k+m+2}\sum_{b = 0}^{2k-m-2} \\
    &\sum_{k' = \text{max}(0,m-2)}^{\ell-2}\sum_{a' = 0}^{2\ell-2k'-m-2}\sum_{b' = 0}^{2k'+m+2}(-1)^{a+a'}G(\ell_0,m,k,a,b;\ell,m,k',a',b')\delta_{p,0}, \\ \label{eq:coep}
    \nn C_+(\ell_+,\ell,m) :=& v_+ F(\ell_+,m+1;\ell,m)\sum_{k = \text{max}(0,m+3)}^{\ell_++2}\sum_{a = 0}^{2\ell_+-2k+m+3}\sum_{b = 0}^{2k-m-3} \\
    \nn &\sum_{k' = \text{max}(0,m-2)}^{\ell-2}\sum_{a' = 0}^{2\ell-2k'-m-2}\sum_{b' = 0}^{2k'+m+2}(-1)^{a+a'}G(\ell_+,m+1,k,a,b;\ell,m,k',a',b') \\
    &\left[\frac{A(\ell_+,m+1)}{p(p^2-4)}\delta_{|\text{mod}(p,2)|,1} + \frac{\pi}{4}(m+1)(\delta_{p,2} - \delta_{p,-2})\right], \\ \label{eq:coem}
    \nn C_-(\ell_-,\ell,m) :=& v_- F(\ell_-,m-1;\ell,m)\sum_{k = \text{max}(0,m+1)}^{\ell_-+2}\sum_{a = 0}^{2\ell_--2k+m+1}\sum_{b = 0}^{2k-m-1} \\
    \nn &\sum_{k' = \text{max}(0,m-2)}^{\ell-2}\sum_{a' = 0}^{2\ell-2k'-m-2}\sum_{b' = 0}^{2k'+m+2}(-1)^{a+a'}G(\ell_-,m-1,k,a,b;\ell,m,k',a',b') \\
    &\left[\frac{A(\ell_-,m-1)}{p(p^2-4)}\delta_{|\text{mod}(p,2)|,1} + \frac{\pi}{4}(m-1)(\delta_{p,2} - \delta_{p,-2})\right],
\end{align}
\end{widetext}
where $v_\pm := v_x \pm iv_y$ is a combination of the components of the velocity in the orbital plane of the binary and $p := \ell'+\ell-a'-a-b'-b$ (here $\ell'$ stands for $\ell_0$, $\ell_+$ or $\ell_-$). Note that in \eqs{eq:coep}{eq:coem} the $\delta_{|\text{mod}(p,2)|,1}$ vanishes in the cases where $p(p^2-4)$ has a root and hence these coefficients are always well defined.

Further, we define the two following functions for short cut:
\begin{widetext}
\begin{align}
    F(\ell',m';\ell,m) :=& \sqrt{\frac{(\ell'+m')!(\ell'-m')!(2\ell'+1)}{(\ell'-2)!(\ell'+2)!}}\sqrt{\frac{(\ell-m)!(\ell+m)!(2\ell+1)}{(\ell+2)!(\ell-2)!}}, \\
    \nn G(\ell',m',k',a',b';\ell,m,k,a,b) :=& \left(\begin{array}{c} \ell'+2 \\ k' \end{array}\right)\left(\begin{array}{c} \ell'-2 \\ k'-m'-2 \end{array}\right)\left(\begin{array}{c} 2\ell'-2k'+m'+2 \\ a' \end{array}\right)\left(\begin{array}{c} 2k'-m'-2 \\ b' \end{array}\right) \\
    & \left(\begin{array}{c} \ell-2 \\ k \end{array}\right)\left(\begin{array}{c} \ell+2 \\ k+m+2 \end{array}\right)\left(\begin{array}{c} 2\ell-2k-m-2 \\ a \end{array}\right)\left(\begin{array}{c} 2k+m+2 \\ b \end{array}\right).
\end{align}
\end{widetext}
%

%The bibliography



\begin{thebibliography}{64}%
\makeatletter
\providecommand \@ifxundefined [1]{%
 \@ifx{#1\undefined}
}%
\providecommand \@ifnum [1]{%
 \ifnum #1\expandafter \@firstoftwo
 \else \expandafter \@secondoftwo
 \fi
}%
\providecommand \@ifx [1]{%
 \ifx #1\expandafter \@firstoftwo
 \else \expandafter \@secondoftwo
 \fi
}%
\providecommand \natexlab [1]{#1}%
\providecommand \enquote  [1]{``#1''}%
\providecommand \bibnamefont  [1]{#1}%
\providecommand \bibfnamefont [1]{#1}%
\providecommand \citenamefont [1]{#1}%
\providecommand \href@noop [0]{\@secondoftwo}%
\providecommand \href [0]{\begingroup \@sanitize@url \@href}%
\providecommand \@href[1]{\@@startlink{#1}\@@href}%
\providecommand \@@href[1]{\endgroup#1\@@endlink}%
\providecommand \@sanitize@url [0]{\catcode `\\12\catcode `\$12\catcode
  `\&12\catcode `\#12\catcode `\^12\catcode `\_12\catcode `\%12\relax}%
\providecommand \@@startlink[1]{}%
\providecommand \@@endlink[0]{}%
\providecommand \url  [0]{\begingroup\@sanitize@url \@url }%
\providecommand \@url [1]{\endgroup\@href {#1}{\urlprefix }}%
\providecommand \urlprefix  [0]{URL }%
\providecommand \Eprint [0]{\href }%
\providecommand \doibase [0]{https://doi.org/}%
\providecommand \selectlanguage [0]{\@gobble}%
\providecommand \bibinfo  [0]{\@secondoftwo}%
\providecommand \bibfield  [0]{\@secondoftwo}%
\providecommand \translation [1]{[#1]}%
\providecommand \BibitemOpen [0]{}%
\providecommand \bibitemStop [0]{}%
\providecommand \bibitemNoStop [0]{.\EOS\space}%
\providecommand \EOS [0]{\spacefactor3000\relax}%
\providecommand \BibitemShut  [1]{\csname bibitem#1\endcsname}%
\let\auto@bib@innerbib\@empty
%</preamble>
\bibitem [{\citenamefont {{The LIGO Scientific Collaboration}}\ and\
  \citenamefont {{the Virgo Collaboration}}(2020{\natexlab{a}})}]{ligo_2020a}%
  \BibitemOpen
  \bibfield  {author} {\bibinfo {author} {\bibnamefont {{The LIGO Scientific
  Collaboration}}}\ and\ \bibinfo {author} {\bibnamefont {{the Virgo
  Collaboration}}},\ }\href {https://doi.org/10.1103/PhysRevD.102.043015}
  {\bibfield  {journal} {\bibinfo  {journal} {\prd}\ }\textbf {\bibinfo
  {volume} {102}},\ \bibinfo {eid} {043015} (\bibinfo {year}
  {2020}{\natexlab{a}})}\BibitemShut {NoStop}%
\bibitem [{\citenamefont {{The LIGO Scientific Collaboration}}\ and\
  \citenamefont {{the Virgo Collaboration}}(2020{\natexlab{b}})}]{ligo_2020b}%
  \BibitemOpen
  \bibfield  {author} {\bibinfo {author} {\bibnamefont {{The LIGO Scientific
  Collaboration}}}\ and\ \bibinfo {author} {\bibnamefont {{the Virgo
  Collaboration}}},\ }\href {https://doi.org/10.3847/2041-8213/ab960f}
  {\bibfield  {journal} {\bibinfo  {journal} {\apjl}\ }\textbf {\bibinfo
  {volume} {896}},\ \bibinfo {eid} {L44} (\bibinfo {year}
  {2020}{\natexlab{b}})},\ \Eprint {https://arxiv.org/abs/2006.12611}
  {arXiv:2006.12611 [astro-ph.HE]} \BibitemShut {NoStop}%
\bibitem [{\citenamefont {Khan}\ \emph {et~al.}(2020)\citenamefont {Khan},
  \citenamefont {Ohme}, \citenamefont {Chatziioannou},\ and\ \citenamefont
  {Hannam}}]{khan_ohme_2020}%
  \BibitemOpen
  \bibfield  {author} {\bibinfo {author} {\bibfnamefont {S.}~\bibnamefont
  {Khan}}, \bibinfo {author} {\bibfnamefont {F.}~\bibnamefont {Ohme}}, \bibinfo
  {author} {\bibfnamefont {K.}~\bibnamefont {Chatziioannou}},\ and\ \bibinfo
  {author} {\bibfnamefont {M.}~\bibnamefont {Hannam}},\ }\href
  {https://doi.org/10.1103/PhysRevD.101.024056} {\bibfield  {journal} {\bibinfo
   {journal} {Phys. Rev. D}\ }\textbf {\bibinfo {volume} {101}},\ \bibinfo
  {pages} {024056} (\bibinfo {year} {2020})}\BibitemShut {NoStop}%
\bibitem [{\citenamefont {Khan}\ \emph {et~al.}(2019)\citenamefont {Khan},
  \citenamefont {Chatziioannou}, \citenamefont {Hannam},\ and\ \citenamefont
  {Ohme}}]{khan_chatziioannou_2019}%
  \BibitemOpen
  \bibfield  {author} {\bibinfo {author} {\bibfnamefont {S.}~\bibnamefont
  {Khan}}, \bibinfo {author} {\bibfnamefont {K.}~\bibnamefont {Chatziioannou}},
  \bibinfo {author} {\bibfnamefont {M.}~\bibnamefont {Hannam}},\ and\ \bibinfo
  {author} {\bibfnamefont {F.}~\bibnamefont {Ohme}},\ }\href
  {https://doi.org/10.1103/PhysRevD.100.024059} {\bibfield  {journal} {\bibinfo
   {journal} {Phys. Rev. D}\ }\textbf {\bibinfo {volume} {100}},\ \bibinfo
  {pages} {024059} (\bibinfo {year} {2019})}\BibitemShut {NoStop}%
\bibitem [{\citenamefont {{Ossokine}}\ \emph {et~al.}(2020)\citenamefont
  {{Ossokine}}, \citenamefont {{Buonanno}}, \citenamefont {{Marsat}},
  \citenamefont {{Cotesta}}, \citenamefont {{Babak}}, \citenamefont
  {{Dietrich}}, \citenamefont {{Haas}}, \citenamefont {{Hinder}}, \citenamefont
  {{Pfeiffer}}, \citenamefont {{P{\"u}rrer}}, \citenamefont {{Woodford}},
  \citenamefont {{Boyle}}, \citenamefont {{Kidder}}, \citenamefont {{Scheel}},\
  and\ \citenamefont {{Szil{\'a}gyi}}}]{ossokine_buonanno_2020}%
  \BibitemOpen
  \bibfield  {author} {\bibinfo {author} {\bibfnamefont {S.}~\bibnamefont
  {{Ossokine}}}, \bibinfo {author} {\bibfnamefont {A.}~\bibnamefont
  {{Buonanno}}}, \bibinfo {author} {\bibfnamefont {S.}~\bibnamefont
  {{Marsat}}}, \bibinfo {author} {\bibfnamefont {R.}~\bibnamefont {{Cotesta}}},
  \bibinfo {author} {\bibfnamefont {S.}~\bibnamefont {{Babak}}}, \bibinfo
  {author} {\bibfnamefont {T.}~\bibnamefont {{Dietrich}}}, \bibinfo {author}
  {\bibfnamefont {R.}~\bibnamefont {{Haas}}}, \bibinfo {author} {\bibfnamefont
  {I.}~\bibnamefont {{Hinder}}}, \bibinfo {author} {\bibfnamefont {H.~P.}\
  \bibnamefont {{Pfeiffer}}}, \bibinfo {author} {\bibfnamefont
  {M.}~\bibnamefont {{P{\"u}rrer}}}, \bibinfo {author} {\bibfnamefont {C.~J.}\
  \bibnamefont {{Woodford}}}, \bibinfo {author} {\bibfnamefont
  {M.}~\bibnamefont {{Boyle}}}, \bibinfo {author} {\bibfnamefont {L.~E.}\
  \bibnamefont {{Kidder}}}, \bibinfo {author} {\bibfnamefont {M.~A.}\
  \bibnamefont {{Scheel}}},\ and\ \bibinfo {author} {\bibfnamefont
  {B.}~\bibnamefont {{Szil{\'a}gyi}}},\ }\href
  {https://doi.org/10.1103/PhysRevD.102.044055} {\bibfield  {journal} {\bibinfo
   {journal} {\prd}\ }\textbf {\bibinfo {volume} {102}},\ \bibinfo {eid}
  {044055} (\bibinfo {year} {2020})},\ \Eprint
  {https://arxiv.org/abs/2004.09442} {arXiv:2004.09442 [gr-qc]} \BibitemShut
  {NoStop}%
\bibitem [{\citenamefont {Pan}\ \emph {et~al.}(2014)\citenamefont {Pan},
  \citenamefont {Buonanno}, \citenamefont {Taracchini}, \citenamefont {Kidder},
  \citenamefont {Mrou\'e}, \citenamefont {Pfeiffer}, \citenamefont {Scheel},\
  and\ \citenamefont {Szil\'agyi}}]{pan_buonanno_2014}%
  \BibitemOpen
  \bibfield  {author} {\bibinfo {author} {\bibfnamefont {Y.}~\bibnamefont
  {Pan}}, \bibinfo {author} {\bibfnamefont {A.}~\bibnamefont {Buonanno}},
  \bibinfo {author} {\bibfnamefont {A.}~\bibnamefont {Taracchini}}, \bibinfo
  {author} {\bibfnamefont {L.~E.}\ \bibnamefont {Kidder}}, \bibinfo {author}
  {\bibfnamefont {A.~H.}\ \bibnamefont {Mrou\'e}}, \bibinfo {author}
  {\bibfnamefont {H.~P.}\ \bibnamefont {Pfeiffer}}, \bibinfo {author}
  {\bibfnamefont {M.~A.}\ \bibnamefont {Scheel}},\ and\ \bibinfo {author}
  {\bibfnamefont {B.}~\bibnamefont {Szil\'agyi}},\ }\href
  {https://doi.org/10.1103/PhysRevD.89.084006} {\bibfield  {journal} {\bibinfo
  {journal} {Phys. Rev. D}\ }\textbf {\bibinfo {volume} {89}},\ \bibinfo
  {pages} {084006} (\bibinfo {year} {2014})}\BibitemShut {NoStop}%
\bibitem [{\citenamefont {Babak}\ \emph {et~al.}(2017)\citenamefont {Babak},
  \citenamefont {Taracchini},\ and\ \citenamefont
  {Buonanno}}]{babak_taracchini_2017}%
  \BibitemOpen
  \bibfield  {author} {\bibinfo {author} {\bibfnamefont {S.}~\bibnamefont
  {Babak}}, \bibinfo {author} {\bibfnamefont {A.}~\bibnamefont {Taracchini}},\
  and\ \bibinfo {author} {\bibfnamefont {A.}~\bibnamefont {Buonanno}},\ }\href
  {https://doi.org/10.1103/PhysRevD.95.024010} {\bibfield  {journal} {\bibinfo
  {journal} {Phys. Rev. D}\ }\textbf {\bibinfo {volume} {95}},\ \bibinfo
  {pages} {024010} (\bibinfo {year} {2017})}\BibitemShut {NoStop}%
\bibitem [{\citenamefont {{Varma}}\ \emph {et~al.}(2019)\citenamefont
  {{Varma}}, \citenamefont {{Field}}, \citenamefont {{Scheel}}, \citenamefont
  {{Blackman}}, \citenamefont {{Gerosa}}, \citenamefont {{Stein}},
  \citenamefont {{Kidder}},\ and\ \citenamefont
  {{Pfeiffer}}}]{varma_field_2019}%
  \BibitemOpen
  \bibfield  {author} {\bibinfo {author} {\bibfnamefont {V.}~\bibnamefont
  {{Varma}}}, \bibinfo {author} {\bibfnamefont {S.~E.}\ \bibnamefont
  {{Field}}}, \bibinfo {author} {\bibfnamefont {M.~A.}\ \bibnamefont
  {{Scheel}}}, \bibinfo {author} {\bibfnamefont {J.}~\bibnamefont
  {{Blackman}}}, \bibinfo {author} {\bibfnamefont {D.}~\bibnamefont
  {{Gerosa}}}, \bibinfo {author} {\bibfnamefont {L.~C.}\ \bibnamefont
  {{Stein}}}, \bibinfo {author} {\bibfnamefont {L.~E.}\ \bibnamefont
  {{Kidder}}},\ and\ \bibinfo {author} {\bibfnamefont {H.~P.}\ \bibnamefont
  {{Pfeiffer}}},\ }\href {https://doi.org/10.1103/PhysRevResearch.1.033015}
  {\bibfield  {journal} {\bibinfo  {journal} {Physical Review Research}\
  }\textbf {\bibinfo {volume} {1}},\ \bibinfo {eid} {033015} (\bibinfo {year}
  {2019})},\ \Eprint {https://arxiv.org/abs/1905.09300} {arXiv:1905.09300
  [gr-qc]} \BibitemShut {NoStop}%
\bibitem [{\citenamefont {{Punturo}}\ and\ \citenamefont {{et
  al.}}(2010)}]{et_2010}%
  \BibitemOpen
  \bibfield  {author} {\bibinfo {author} {\bibfnamefont {M.}~\bibnamefont
  {{Punturo}}}\ and\ \bibinfo {author} {\bibnamefont {{et al.}}},\ }\href
  {https://doi.org/10.1088/0264-9381/27/19/194002} {\bibfield  {journal}
  {\bibinfo  {journal} {Classical and Quantum Gravity}\ }\textbf {\bibinfo
  {volume} {27}},\ \bibinfo {eid} {194002} (\bibinfo {year}
  {2010})}\BibitemShut {NoStop}%
\bibitem [{\citenamefont {{Reitze}}\ \emph {et~al.}(2019)\citenamefont
  {{Reitze}}, \citenamefont {{Adhikari}}, \citenamefont {{Ballmer}},
  \citenamefont {{Barish}}, \citenamefont {{Barsotti}}, \citenamefont
  {{Billingsley}}, \citenamefont {{Brown}}, \citenamefont {{Chen}},
  \citenamefont {{Coyne}}, \citenamefont {{Eisenstein}}, \citenamefont
  {{Evans}}, \citenamefont {{Fritschel}}, \citenamefont {{Hall}}, \citenamefont
  {{Lazzarini}}, \citenamefont {{Lovelace}}, \citenamefont {{Read}},
  \citenamefont {{Sathyaprakash}}, \citenamefont {{Shoemaker}}, \citenamefont
  {{Smith}}, \citenamefont {{Torrie}}, \citenamefont {{Vitale}}, \citenamefont
  {{Weiss}}, \citenamefont {{Wipf}},\ and\ \citenamefont
  {{Zucker}}}]{cosmic_explorer_2019}%
  \BibitemOpen
  \bibfield  {author} {\bibinfo {author} {\bibfnamefont {D.}~\bibnamefont
  {{Reitze}}}, \bibinfo {author} {\bibfnamefont {R.~X.}\ \bibnamefont
  {{Adhikari}}}, \bibinfo {author} {\bibfnamefont {S.}~\bibnamefont
  {{Ballmer}}}, \bibinfo {author} {\bibfnamefont {B.}~\bibnamefont {{Barish}}},
  \bibinfo {author} {\bibfnamefont {L.}~\bibnamefont {{Barsotti}}}, \bibinfo
  {author} {\bibfnamefont {G.}~\bibnamefont {{Billingsley}}}, \bibinfo {author}
  {\bibfnamefont {D.~A.}\ \bibnamefont {{Brown}}}, \bibinfo {author}
  {\bibfnamefont {Y.}~\bibnamefont {{Chen}}}, \bibinfo {author} {\bibfnamefont
  {D.}~\bibnamefont {{Coyne}}}, \bibinfo {author} {\bibfnamefont
  {R.}~\bibnamefont {{Eisenstein}}}, \bibinfo {author} {\bibfnamefont
  {M.}~\bibnamefont {{Evans}}}, \bibinfo {author} {\bibfnamefont
  {P.}~\bibnamefont {{Fritschel}}}, \bibinfo {author} {\bibfnamefont {E.~D.}\
  \bibnamefont {{Hall}}}, \bibinfo {author} {\bibfnamefont {A.}~\bibnamefont
  {{Lazzarini}}}, \bibinfo {author} {\bibfnamefont {G.}~\bibnamefont
  {{Lovelace}}}, \bibinfo {author} {\bibfnamefont {J.}~\bibnamefont {{Read}}},
  \bibinfo {author} {\bibfnamefont {B.~S.}\ \bibnamefont {{Sathyaprakash}}},
  \bibinfo {author} {\bibfnamefont {D.}~\bibnamefont {{Shoemaker}}}, \bibinfo
  {author} {\bibfnamefont {J.}~\bibnamefont {{Smith}}}, \bibinfo {author}
  {\bibfnamefont {C.}~\bibnamefont {{Torrie}}}, \bibinfo {author}
  {\bibfnamefont {S.}~\bibnamefont {{Vitale}}}, \bibinfo {author}
  {\bibfnamefont {R.}~\bibnamefont {{Weiss}}}, \bibinfo {author} {\bibfnamefont
  {C.}~\bibnamefont {{Wipf}}},\ and\ \bibinfo {author} {\bibfnamefont
  {M.}~\bibnamefont {{Zucker}}},\ }in\ \href@noop {} {\emph {\bibinfo
  {booktitle} {\baas}}},\ Vol.~\bibinfo {volume} {51}\ (\bibinfo {year}
  {2019})\ p.~\bibinfo {pages} {35}\BibitemShut {NoStop}%
\bibitem [{\citenamefont {{Amaro-Seoane }}\ \emph {et~al.}(2017)\citenamefont
  {{Amaro-Seoane }} \emph {et~al.}}]{lisa_2017}%
  \BibitemOpen
  \bibfield  {author} {\bibinfo {author} {\bibfnamefont {P.}~\bibnamefont
  {{Amaro-Seoane }}} \emph {et~al.},\ }\href@noop {} {\bibfield  {journal}
  {\bibinfo  {journal} {ArXiv e-prints}\ } (\bibinfo {year} {2017})},\ \Eprint
  {https://arxiv.org/abs/1702.00786} {arXiv:1702.00786 [astro-ph.IM]}
  \BibitemShut {NoStop}%
\bibitem [{\citenamefont {{Luo}}\ and\ \citenamefont {{et
  al.}}(2016)}]{tianqin_2016}%
  \BibitemOpen
  \bibfield  {author} {\bibinfo {author} {\bibfnamefont {J.}~\bibnamefont
  {{Luo}}}\ and\ \bibinfo {author} {\bibnamefont {{et al.}}},\ }\href
  {https://doi.org/10.1088/0264-9381/33/3/035010} {\bibfield  {journal}
  {\bibinfo  {journal} {Classical and Quantum Gravity}\ }\textbf {\bibinfo
  {volume} {33}},\ \bibinfo {eid} {035010} (\bibinfo {year} {2016})},\ \Eprint
  {https://arxiv.org/abs/1512.02076} {arXiv:1512.02076 [astro-ph.IM]}
  \BibitemShut {NoStop}%
\bibitem [{\citenamefont {{Gong}}\ \emph {et~al.}(2015)\citenamefont {{Gong}},
  \citenamefont {{Lau}}, \citenamefont {{Xu}}, \citenamefont {{Amaro-Seoane}},
  \citenamefont {{Bai}}, \citenamefont {{Bian}}, \citenamefont {{Cao}},
  \citenamefont {{Chen}}, \citenamefont {{Chen}}, \citenamefont {{Ding}},
  \citenamefont {{Dong}}, \citenamefont {{Gao}}, \citenamefont {{Heinzel}},
  \citenamefont {{Li}}, \citenamefont {{Li}}, \citenamefont {{Liu}},
  \citenamefont {{Luo}}, \citenamefont {{Shao}}, \citenamefont {{Spurzem}},
  \citenamefont {{Sun}}, \citenamefont {{Tang}}, \citenamefont {{Wang}},
  \citenamefont {{Xu}}, \citenamefont {{Yu}}, \citenamefont {{Yuan}},
  \citenamefont {{Zhang}},\ and\ \citenamefont {{Zhou}}}]{taiji_2015}%
  \BibitemOpen
  \bibfield  {author} {\bibinfo {author} {\bibfnamefont {X.}~\bibnamefont
  {{Gong}}}, \bibinfo {author} {\bibfnamefont {Y.-K.}\ \bibnamefont {{Lau}}},
  \bibinfo {author} {\bibfnamefont {S.}~\bibnamefont {{Xu}}}, \bibinfo {author}
  {\bibfnamefont {P.}~\bibnamefont {{Amaro-Seoane}}}, \bibinfo {author}
  {\bibfnamefont {S.}~\bibnamefont {{Bai}}}, \bibinfo {author} {\bibfnamefont
  {X.}~\bibnamefont {{Bian}}}, \bibinfo {author} {\bibfnamefont
  {Z.}~\bibnamefont {{Cao}}}, \bibinfo {author} {\bibfnamefont
  {G.}~\bibnamefont {{Chen}}}, \bibinfo {author} {\bibfnamefont
  {X.}~\bibnamefont {{Chen}}}, \bibinfo {author} {\bibfnamefont
  {Y.}~\bibnamefont {{Ding}}}, \bibinfo {author} {\bibfnamefont
  {P.}~\bibnamefont {{Dong}}}, \bibinfo {author} {\bibfnamefont
  {W.}~\bibnamefont {{Gao}}}, \bibinfo {author} {\bibfnamefont
  {G.}~\bibnamefont {{Heinzel}}}, \bibinfo {author} {\bibfnamefont
  {M.}~\bibnamefont {{Li}}}, \bibinfo {author} {\bibfnamefont {S.}~\bibnamefont
  {{Li}}}, \bibinfo {author} {\bibfnamefont {F.}~\bibnamefont {{Liu}}},
  \bibinfo {author} {\bibfnamefont {Z.}~\bibnamefont {{Luo}}}, \bibinfo
  {author} {\bibfnamefont {M.}~\bibnamefont {{Shao}}}, \bibinfo {author}
  {\bibfnamefont {R.}~\bibnamefont {{Spurzem}}}, \bibinfo {author}
  {\bibfnamefont {B.}~\bibnamefont {{Sun}}}, \bibinfo {author} {\bibfnamefont
  {W.}~\bibnamefont {{Tang}}}, \bibinfo {author} {\bibfnamefont
  {Y.}~\bibnamefont {{Wang}}}, \bibinfo {author} {\bibfnamefont
  {P.}~\bibnamefont {{Xu}}}, \bibinfo {author} {\bibfnamefont {P.}~\bibnamefont
  {{Yu}}}, \bibinfo {author} {\bibfnamefont {Y.}~\bibnamefont {{Yuan}}},
  \bibinfo {author} {\bibfnamefont {X.}~\bibnamefont {{Zhang}}},\ and\ \bibinfo
  {author} {\bibfnamefont {Z.}~\bibnamefont {{Zhou}}},\ }in\ \href
  {https://doi.org/10.1088/1742-6596/610/1/012011} {\emph {\bibinfo {booktitle}
  {Journal of Physics Conference Series}}},\ \bibinfo {series} {Journal of
  Physics Conference Series}, Vol.\ \bibinfo {volume} {610}\ (\bibinfo {year}
  {2015})\ p.\ \bibinfo {pages} {012011},\ \Eprint
  {https://arxiv.org/abs/1410.7296} {arXiv:1410.7296 [gr-qc]} \BibitemShut
  {NoStop}%
\bibitem [{\citenamefont {{Blanchet}}(2006)}]{blanchet_2006}%
  \BibitemOpen
  \bibfield  {author} {\bibinfo {author} {\bibfnamefont {L.}~\bibnamefont
  {{Blanchet}}},\ }\href {https://doi.org/10.12942/lrr-2006-4} {\bibfield
  {journal} {\bibinfo  {journal} {Living Reviews in Relativity}\ }\textbf
  {\bibinfo {volume} {9}},\ \bibinfo {eid} {4} (\bibinfo {year}
  {2006})}\BibitemShut {NoStop}%
\bibitem [{\citenamefont {{Santamar{\'{\i}}a}}\ \emph
  {et~al.}(2010)\citenamefont {{Santamar{\'{\i}}a}}, \citenamefont {{Ohme}},
  \citenamefont {{Ajith}}, \citenamefont {{Br{\"u}gmann}}, \citenamefont
  {{Dorband}}, \citenamefont {{Hannam}}, \citenamefont {{Husa}}, \citenamefont
  {{M{\"o}sta}}, \citenamefont {{Pollney}}, \citenamefont {{Reisswig}},
  \citenamefont {{Robinson}}, \citenamefont {{Seiler}},\ and\ \citenamefont
  {{Krishnan}}}]{santamaria_ohme_2010}%
  \BibitemOpen
  \bibfield  {author} {\bibinfo {author} {\bibfnamefont {L.}~\bibnamefont
  {{Santamar{\'{\i}}a}}}, \bibinfo {author} {\bibfnamefont {F.}~\bibnamefont
  {{Ohme}}}, \bibinfo {author} {\bibfnamefont {P.}~\bibnamefont {{Ajith}}},
  \bibinfo {author} {\bibfnamefont {B.}~\bibnamefont {{Br{\"u}gmann}}},
  \bibinfo {author} {\bibfnamefont {N.}~\bibnamefont {{Dorband}}}, \bibinfo
  {author} {\bibfnamefont {M.}~\bibnamefont {{Hannam}}}, \bibinfo {author}
  {\bibfnamefont {S.}~\bibnamefont {{Husa}}}, \bibinfo {author} {\bibfnamefont
  {P.}~\bibnamefont {{M{\"o}sta}}}, \bibinfo {author} {\bibfnamefont
  {D.}~\bibnamefont {{Pollney}}}, \bibinfo {author} {\bibfnamefont
  {C.}~\bibnamefont {{Reisswig}}}, \bibinfo {author} {\bibfnamefont {E.~L.}\
  \bibnamefont {{Robinson}}}, \bibinfo {author} {\bibfnamefont
  {J.}~\bibnamefont {{Seiler}}},\ and\ \bibinfo {author} {\bibfnamefont
  {B.}~\bibnamefont {{Krishnan}}},\ }\href
  {https://doi.org/10.1103/PhysRevD.82.064016} {\bibfield  {journal} {\bibinfo
  {journal} {Phys Rev D}\ }\textbf {\bibinfo {volume} {82}},\ \bibinfo {pages}
  {064016} (\bibinfo {year} {2010})},\ \Eprint
  {https://arxiv.org/abs/1005.3306} {arXiv:1005.3306 [gr-qc]} \BibitemShut
  {NoStop}%
\bibitem [{\citenamefont {{Hannam}}\ \emph {et~al.}(2014)\citenamefont
  {{Hannam}}, \citenamefont {{Schmidt}}, \citenamefont {{Boh{\'e}}},
  \citenamefont {{Haegel}}, \citenamefont {{Husa}}, \citenamefont {{Ohme}},
  \citenamefont {{Pratten}},\ and\ \citenamefont
  {{P{\"u}rrer}}}]{hannam_schmidt_2014}%
  \BibitemOpen
  \bibfield  {author} {\bibinfo {author} {\bibfnamefont {M.}~\bibnamefont
  {{Hannam}}}, \bibinfo {author} {\bibfnamefont {P.}~\bibnamefont {{Schmidt}}},
  \bibinfo {author} {\bibfnamefont {A.}~\bibnamefont {{Boh{\'e}}}}, \bibinfo
  {author} {\bibfnamefont {L.}~\bibnamefont {{Haegel}}}, \bibinfo {author}
  {\bibfnamefont {S.}~\bibnamefont {{Husa}}}, \bibinfo {author} {\bibfnamefont
  {F.}~\bibnamefont {{Ohme}}}, \bibinfo {author} {\bibfnamefont
  {G.}~\bibnamefont {{Pratten}}},\ and\ \bibinfo {author} {\bibfnamefont
  {M.}~\bibnamefont {{P{\"u}rrer}}},\ }\href
  {https://doi.org/10.1103/PhysRevLett.113.151101} {\bibfield  {journal}
  {\bibinfo  {journal} {\prl}\ }\textbf {\bibinfo {volume} {113}},\ \bibinfo
  {eid} {151101} (\bibinfo {year} {2014})},\ \Eprint
  {https://arxiv.org/abs/1308.3271} {arXiv:1308.3271 [gr-qc]} \BibitemShut
  {NoStop}%
\bibitem [{\citenamefont {{Buonanno}}\ and\ \citenamefont
  {{Damour}}(1999)}]{buonanno_damour_1999}%
  \BibitemOpen
  \bibfield  {author} {\bibinfo {author} {\bibfnamefont {A.}~\bibnamefont
  {{Buonanno}}}\ and\ \bibinfo {author} {\bibfnamefont {T.}~\bibnamefont
  {{Damour}}},\ }\href {https://doi.org/10.1103/PhysRevD.59.084006} {\bibfield
  {journal} {\bibinfo  {journal} {\prd}\ }\textbf {\bibinfo {volume} {59}},\
  \bibinfo {eid} {084006} (\bibinfo {year} {1999})},\ \Eprint
  {https://arxiv.org/abs/gr-qc/9811091} {gr-qc/9811091} \BibitemShut {NoStop}%
\bibitem [{\citenamefont {{Boyle}}\ \emph {et~al.}(2019)\citenamefont
  {{Boyle}}, \citenamefont {{Hemberger}}, \citenamefont {{Iozzo}},
  \citenamefont {{Lovelace}}, \citenamefont {{Ossokine}}, \citenamefont
  {{Pfeiffer}}, \citenamefont {{Scheel}}, \citenamefont {{Stein}},
  \citenamefont {{Woodford}}, \citenamefont {{Zimmerman}}, \citenamefont
  {{Afshari}}, \citenamefont {{Barkett}}, \citenamefont {{Blackman}},
  \citenamefont {{Chatziioannou}}, \citenamefont {{Chu}}, \citenamefont
  {{Demos}}, \citenamefont {{Deppe}}, \citenamefont {{Field}}, \citenamefont
  {{Fischer}}, \citenamefont {{Foley}}, \citenamefont {{Fong}}, \citenamefont
  {{Garcia}}, \citenamefont {{Giesler}}, \citenamefont {{Hebert}},
  \citenamefont {{Hinder}}, \citenamefont {{Katebi}}, \citenamefont {{Khan}},
  \citenamefont {{Kidder}}, \citenamefont {{Kumar}}, \citenamefont {{Kuper}},
  \citenamefont {{Lim}}, \citenamefont {{Okounkova}}, \citenamefont
  {{Ramirez}}, \citenamefont {{Rodriguez}}, \citenamefont {{R{\"u}ter}},
  \citenamefont {{Schmidt}}, \citenamefont {{Szilagyi}}, \citenamefont
  {{Teukolsky}}, \citenamefont {{Varma}},\ and\ \citenamefont
  {{Walker}}}]{SXS_2019}%
  \BibitemOpen
  \bibfield  {author} {\bibinfo {author} {\bibfnamefont {M.}~\bibnamefont
  {{Boyle}}}, \bibinfo {author} {\bibfnamefont {D.}~\bibnamefont
  {{Hemberger}}}, \bibinfo {author} {\bibfnamefont {D.~A.~B.}\ \bibnamefont
  {{Iozzo}}}, \bibinfo {author} {\bibfnamefont {G.}~\bibnamefont {{Lovelace}}},
  \bibinfo {author} {\bibfnamefont {S.}~\bibnamefont {{Ossokine}}}, \bibinfo
  {author} {\bibfnamefont {H.~P.}\ \bibnamefont {{Pfeiffer}}}, \bibinfo
  {author} {\bibfnamefont {M.~A.}\ \bibnamefont {{Scheel}}}, \bibinfo {author}
  {\bibfnamefont {L.~C.}\ \bibnamefont {{Stein}}}, \bibinfo {author}
  {\bibfnamefont {C.~J.}\ \bibnamefont {{Woodford}}}, \bibinfo {author}
  {\bibfnamefont {A.~B.}\ \bibnamefont {{Zimmerman}}}, \bibinfo {author}
  {\bibfnamefont {N.}~\bibnamefont {{Afshari}}}, \bibinfo {author}
  {\bibfnamefont {K.}~\bibnamefont {{Barkett}}}, \bibinfo {author}
  {\bibfnamefont {J.}~\bibnamefont {{Blackman}}}, \bibinfo {author}
  {\bibfnamefont {K.}~\bibnamefont {{Chatziioannou}}}, \bibinfo {author}
  {\bibfnamefont {T.}~\bibnamefont {{Chu}}}, \bibinfo {author} {\bibfnamefont
  {N.}~\bibnamefont {{Demos}}}, \bibinfo {author} {\bibfnamefont
  {N.}~\bibnamefont {{Deppe}}}, \bibinfo {author} {\bibfnamefont {S.~E.}\
  \bibnamefont {{Field}}}, \bibinfo {author} {\bibfnamefont {N.~L.}\
  \bibnamefont {{Fischer}}}, \bibinfo {author} {\bibfnamefont {E.}~\bibnamefont
  {{Foley}}}, \bibinfo {author} {\bibfnamefont {H.}~\bibnamefont {{Fong}}},
  \bibinfo {author} {\bibfnamefont {A.}~\bibnamefont {{Garcia}}}, \bibinfo
  {author} {\bibfnamefont {M.}~\bibnamefont {{Giesler}}}, \bibinfo {author}
  {\bibfnamefont {F.}~\bibnamefont {{Hebert}}}, \bibinfo {author}
  {\bibfnamefont {I.}~\bibnamefont {{Hinder}}}, \bibinfo {author}
  {\bibfnamefont {R.}~\bibnamefont {{Katebi}}}, \bibinfo {author}
  {\bibfnamefont {H.}~\bibnamefont {{Khan}}}, \bibinfo {author} {\bibfnamefont
  {L.~E.}\ \bibnamefont {{Kidder}}}, \bibinfo {author} {\bibfnamefont
  {P.}~\bibnamefont {{Kumar}}}, \bibinfo {author} {\bibfnamefont
  {K.}~\bibnamefont {{Kuper}}}, \bibinfo {author} {\bibfnamefont
  {H.}~\bibnamefont {{Lim}}}, \bibinfo {author} {\bibfnamefont
  {M.}~\bibnamefont {{Okounkova}}}, \bibinfo {author} {\bibfnamefont
  {T.}~\bibnamefont {{Ramirez}}}, \bibinfo {author} {\bibfnamefont
  {S.}~\bibnamefont {{Rodriguez}}}, \bibinfo {author} {\bibfnamefont {H.~R.}\
  \bibnamefont {{R{\"u}ter}}}, \bibinfo {author} {\bibfnamefont
  {P.}~\bibnamefont {{Schmidt}}}, \bibinfo {author} {\bibfnamefont
  {B.}~\bibnamefont {{Szilagyi}}}, \bibinfo {author} {\bibfnamefont {S.~A.}\
  \bibnamefont {{Teukolsky}}}, \bibinfo {author} {\bibfnamefont
  {V.}~\bibnamefont {{Varma}}},\ and\ \bibinfo {author} {\bibfnamefont
  {M.}~\bibnamefont {{Walker}}},\ }\href
  {https://doi.org/10.1088/1361-6382/ab34e2} {\bibfield  {journal} {\bibinfo
  {journal} {Classical and Quantum Gravity}\ }\textbf {\bibinfo {volume}
  {36}},\ \bibinfo {eid} {195006} (\bibinfo {year} {2019})},\ \Eprint
  {https://arxiv.org/abs/1904.04831} {arXiv:1904.04831 [gr-qc]} \BibitemShut
  {NoStop}%
\bibitem [{\citenamefont {{Healy}}\ \emph {et~al.}(2019)\citenamefont
  {{Healy}}, \citenamefont {{Lousto}}, \citenamefont {{Lange}}, \citenamefont
  {{O'Shaughnessy}}, \citenamefont {{Zlochower}},\ and\ \citenamefont
  {{Campanelli}}}]{RIT_2019}%
  \BibitemOpen
  \bibfield  {author} {\bibinfo {author} {\bibfnamefont {J.}~\bibnamefont
  {{Healy}}}, \bibinfo {author} {\bibfnamefont {C.~O.}\ \bibnamefont
  {{Lousto}}}, \bibinfo {author} {\bibfnamefont {J.}~\bibnamefont {{Lange}}},
  \bibinfo {author} {\bibfnamefont {R.}~\bibnamefont {{O'Shaughnessy}}},
  \bibinfo {author} {\bibfnamefont {Y.}~\bibnamefont {{Zlochower}}},\ and\
  \bibinfo {author} {\bibfnamefont {M.}~\bibnamefont {{Campanelli}}},\ }\href
  {https://doi.org/10.1103/PhysRevD.100.024021} {\bibfield  {journal} {\bibinfo
   {journal} {\prd}\ }\textbf {\bibinfo {volume} {100}},\ \bibinfo {eid}
  {024021} (\bibinfo {year} {2019})},\ \Eprint
  {https://arxiv.org/abs/1901.02553} {arXiv:1901.02553 [gr-qc]} \BibitemShut
  {NoStop}%
\bibitem [{\citenamefont {{Jani}}\ \emph {et~al.}(2016)\citenamefont {{Jani}},
  \citenamefont {{Healy}}, \citenamefont {{Clark}}, \citenamefont {{London}},
  \citenamefont {{Laguna}},\ and\ \citenamefont {{Shoemaker}}}]{GT_2016}%
  \BibitemOpen
  \bibfield  {author} {\bibinfo {author} {\bibfnamefont {K.}~\bibnamefont
  {{Jani}}}, \bibinfo {author} {\bibfnamefont {J.}~\bibnamefont {{Healy}}},
  \bibinfo {author} {\bibfnamefont {J.~A.}\ \bibnamefont {{Clark}}}, \bibinfo
  {author} {\bibfnamefont {L.}~\bibnamefont {{London}}}, \bibinfo {author}
  {\bibfnamefont {P.}~\bibnamefont {{Laguna}}},\ and\ \bibinfo {author}
  {\bibfnamefont {D.}~\bibnamefont {{Shoemaker}}},\ }\href
  {https://doi.org/10.1088/0264-9381/33/20/204001} {\bibfield  {journal}
  {\bibinfo  {journal} {Classical and Quantum Gravity}\ }\textbf {\bibinfo
  {volume} {33}},\ \bibinfo {eid} {204001} (\bibinfo {year} {2016})},\ \Eprint
  {https://arxiv.org/abs/1605.03204} {arXiv:1605.03204 [gr-qc]} \BibitemShut
  {NoStop}%
\bibitem [{\citenamefont {{Blackman}}\ \emph {et~al.}(2017)\citenamefont
  {{Blackman}}, \citenamefont {{Field}}, \citenamefont {{Scheel}},
  \citenamefont {{Galley}}, \citenamefont {{Ott}}, \citenamefont {{Boyle}},
  \citenamefont {{Kidder}}, \citenamefont {{Pfeiffer}},\ and\ \citenamefont
  {{Szil{\'a}gyi}}}]{blackman_field_2017}%
  \BibitemOpen
  \bibfield  {author} {\bibinfo {author} {\bibfnamefont {J.}~\bibnamefont
  {{Blackman}}}, \bibinfo {author} {\bibfnamefont {S.~E.}\ \bibnamefont
  {{Field}}}, \bibinfo {author} {\bibfnamefont {M.~A.}\ \bibnamefont
  {{Scheel}}}, \bibinfo {author} {\bibfnamefont {C.~R.}\ \bibnamefont
  {{Galley}}}, \bibinfo {author} {\bibfnamefont {C.~D.}\ \bibnamefont {{Ott}}},
  \bibinfo {author} {\bibfnamefont {M.}~\bibnamefont {{Boyle}}}, \bibinfo
  {author} {\bibfnamefont {L.~E.}\ \bibnamefont {{Kidder}}}, \bibinfo {author}
  {\bibfnamefont {H.~P.}\ \bibnamefont {{Pfeiffer}}},\ and\ \bibinfo {author}
  {\bibfnamefont {B.}~\bibnamefont {{Szil{\'a}gyi}}},\ }\href
  {https://doi.org/10.1103/PhysRevD.96.024058} {\bibfield  {journal} {\bibinfo
  {journal} {\prd}\ }\textbf {\bibinfo {volume} {96}},\ \bibinfo {eid} {024058}
  (\bibinfo {year} {2017})},\ \Eprint {https://arxiv.org/abs/1705.07089}
  {arXiv:1705.07089 [gr-qc]} \BibitemShut {NoStop}%
\bibitem [{\citenamefont {{Bahcall}}(1988)}]{bahcall_1988}%
  \BibitemOpen
  \bibfield  {author} {\bibinfo {author} {\bibfnamefont {N.~A.}\ \bibnamefont
  {{Bahcall}}},\ }\href {https://doi.org/10.1146/annurev.aa.26.090188.003215}
  {\bibfield  {journal} {\bibinfo  {journal} {\araa}\ }\textbf {\bibinfo
  {volume} {26}},\ \bibinfo {pages} {631} (\bibinfo {year} {1988})}\BibitemShut
  {NoStop}%
\bibitem [{\citenamefont {{Scrimgeour}}\ \emph {et~al.}(2016)\citenamefont
  {{Scrimgeour}}, \citenamefont {{Davis}}, \citenamefont {{Blake}},
  \citenamefont {{Staveley-Smith}}, \citenamefont {{Magoulas}}, \citenamefont
  {{Springob}}, \citenamefont {{Beutler}}, \citenamefont {{Colless}},
  \citenamefont {{Johnson}}, \citenamefont {{Jones}}, \citenamefont {{Koda}},
  \citenamefont {{Lucey}}, \citenamefont {{Ma}}, \citenamefont {{Mould}},\ and\
  \citenamefont {{Poole}}}]{scrimgeour_davis_2016}%
  \BibitemOpen
  \bibfield  {author} {\bibinfo {author} {\bibfnamefont {M.~I.}\ \bibnamefont
  {{Scrimgeour}}}, \bibinfo {author} {\bibfnamefont {T.~M.}\ \bibnamefont
  {{Davis}}}, \bibinfo {author} {\bibfnamefont {C.}~\bibnamefont {{Blake}}},
  \bibinfo {author} {\bibfnamefont {L.}~\bibnamefont {{Staveley-Smith}}},
  \bibinfo {author} {\bibfnamefont {C.}~\bibnamefont {{Magoulas}}}, \bibinfo
  {author} {\bibfnamefont {C.~M.}\ \bibnamefont {{Springob}}}, \bibinfo
  {author} {\bibfnamefont {F.}~\bibnamefont {{Beutler}}}, \bibinfo {author}
  {\bibfnamefont {M.}~\bibnamefont {{Colless}}}, \bibinfo {author}
  {\bibfnamefont {A.}~\bibnamefont {{Johnson}}}, \bibinfo {author}
  {\bibfnamefont {D.~H.}\ \bibnamefont {{Jones}}}, \bibinfo {author}
  {\bibfnamefont {J.}~\bibnamefont {{Koda}}}, \bibinfo {author} {\bibfnamefont
  {J.~R.}\ \bibnamefont {{Lucey}}}, \bibinfo {author} {\bibfnamefont {Y.-Z.}\
  \bibnamefont {{Ma}}}, \bibinfo {author} {\bibfnamefont {J.}~\bibnamefont
  {{Mould}}},\ and\ \bibinfo {author} {\bibfnamefont {G.~B.}\ \bibnamefont
  {{Poole}}},\ }\href {https://doi.org/10.1093/mnras/stv2146} {\bibfield
  {journal} {\bibinfo  {journal} {\mnras}\ }\textbf {\bibinfo {volume} {455}},\
  \bibinfo {pages} {386} (\bibinfo {year} {2016})},\ \Eprint
  {https://arxiv.org/abs/1511.06930} {arXiv:1511.06930 [astro-ph.CO]}
  \BibitemShut {NoStop}%
\bibitem [{\citenamefont {{Colin}}\ \emph {et~al.}(2017)\citenamefont
  {{Colin}}, \citenamefont {{Mohayaee}}, \citenamefont {{Rameez}},\ and\
  \citenamefont {{Sarkar}}}]{colin_mohayaee_2017}%
  \BibitemOpen
  \bibfield  {author} {\bibinfo {author} {\bibfnamefont {J.}~\bibnamefont
  {{Colin}}}, \bibinfo {author} {\bibfnamefont {R.}~\bibnamefont {{Mohayaee}}},
  \bibinfo {author} {\bibfnamefont {M.}~\bibnamefont {{Rameez}}},\ and\
  \bibinfo {author} {\bibfnamefont {S.}~\bibnamefont {{Sarkar}}},\ }\href
  {https://doi.org/10.1093/mnras/stx1631} {\bibfield  {journal} {\bibinfo
  {journal} {\mnras}\ }\textbf {\bibinfo {volume} {471}},\ \bibinfo {pages}
  {1045} (\bibinfo {year} {2017})},\ \Eprint {https://arxiv.org/abs/1703.09376}
  {arXiv:1703.09376 [astro-ph.CO]} \BibitemShut {NoStop}%
\bibitem [{\citenamefont {{Zinn}}\ and\ \citenamefont
  {{West}}(1984)}]{zinn_west_1984}%
  \BibitemOpen
  \bibfield  {author} {\bibinfo {author} {\bibfnamefont {R.}~\bibnamefont
  {{Zinn}}}\ and\ \bibinfo {author} {\bibfnamefont {M.~J.}\ \bibnamefont
  {{West}}},\ }\href {https://doi.org/10.1086/190947} {\bibfield  {journal}
  {\bibinfo  {journal} {\apjs}\ }\textbf {\bibinfo {volume} {55}},\ \bibinfo
  {pages} {45} (\bibinfo {year} {1984})}\BibitemShut {NoStop}%
\bibitem [{\citenamefont {{Girardi}}\ \emph {et~al.}(1996)\citenamefont
  {{Girardi}}, \citenamefont {{Fadda}}, \citenamefont {{Giuricin}},
  \citenamefont {{Mardirossian}}, \citenamefont {{Mezzetti}},\ and\
  \citenamefont {{Biviano}}}]{girardi_fadda_1996}%
  \BibitemOpen
  \bibfield  {author} {\bibinfo {author} {\bibfnamefont {M.}~\bibnamefont
  {{Girardi}}}, \bibinfo {author} {\bibfnamefont {D.}~\bibnamefont {{Fadda}}},
  \bibinfo {author} {\bibfnamefont {G.}~\bibnamefont {{Giuricin}}}, \bibinfo
  {author} {\bibfnamefont {F.}~\bibnamefont {{Mardirossian}}}, \bibinfo
  {author} {\bibfnamefont {M.}~\bibnamefont {{Mezzetti}}},\ and\ \bibinfo
  {author} {\bibfnamefont {A.}~\bibnamefont {{Biviano}}},\ }\href
  {https://doi.org/10.1086/176711} {\bibfield  {journal} {\bibinfo  {journal}
  {\apj}\ }\textbf {\bibinfo {volume} {457}},\ \bibinfo {pages} {61} (\bibinfo
  {year} {1996})},\ \Eprint {https://arxiv.org/abs/astro-ph/9507031}
  {arXiv:astro-ph/9507031 [astro-ph]} \BibitemShut {NoStop}%
\bibitem [{\citenamefont {{Carlberg}}\ \emph {et~al.}(1996)\citenamefont
  {{Carlberg}}, \citenamefont {{Yee}}, \citenamefont {{Ellingson}},
  \citenamefont {{Abraham}}, \citenamefont {{Gravel}}, \citenamefont
  {{Morris}},\ and\ \citenamefont {{Pritchet}}}]{carlberg_yee_1996}%
  \BibitemOpen
  \bibfield  {author} {\bibinfo {author} {\bibfnamefont {R.~G.}\ \bibnamefont
  {{Carlberg}}}, \bibinfo {author} {\bibfnamefont {H.~K.~C.}\ \bibnamefont
  {{Yee}}}, \bibinfo {author} {\bibfnamefont {E.}~\bibnamefont {{Ellingson}}},
  \bibinfo {author} {\bibfnamefont {R.}~\bibnamefont {{Abraham}}}, \bibinfo
  {author} {\bibfnamefont {P.}~\bibnamefont {{Gravel}}}, \bibinfo {author}
  {\bibfnamefont {S.}~\bibnamefont {{Morris}}},\ and\ \bibinfo {author}
  {\bibfnamefont {C.~J.}\ \bibnamefont {{Pritchet}}},\ }\href
  {https://doi.org/10.1086/177125} {\bibfield  {journal} {\bibinfo  {journal}
  {\apj}\ }\textbf {\bibinfo {volume} {462}},\ \bibinfo {pages} {32} (\bibinfo
  {year} {1996})},\ \Eprint {https://arxiv.org/abs/astro-ph/9509034}
  {arXiv:astro-ph/9509034 [astro-ph]} \BibitemShut {NoStop}%
\bibitem [{\citenamefont {{Springel}}\ \emph {et~al.}(2001)\citenamefont
  {{Springel}}, \citenamefont {{White}}, \citenamefont {{Tormen}},\ and\
  \citenamefont {{Kauffmann}}}]{springel_white_2001}%
  \BibitemOpen
  \bibfield  {author} {\bibinfo {author} {\bibfnamefont {V.}~\bibnamefont
  {{Springel}}}, \bibinfo {author} {\bibfnamefont {S.~D.~M.}\ \bibnamefont
  {{White}}}, \bibinfo {author} {\bibfnamefont {G.}~\bibnamefont {{Tormen}}},\
  and\ \bibinfo {author} {\bibfnamefont {G.}~\bibnamefont {{Kauffmann}}},\
  }\href {https://doi.org/10.1046/j.1365-8711.2001.04912.x} {\bibfield
  {journal} {\bibinfo  {journal} {\mnras}\ }\textbf {\bibinfo {volume} {328}},\
  \bibinfo {pages} {726} (\bibinfo {year} {2001})},\ \Eprint
  {https://arxiv.org/abs/astro-ph/0012055} {arXiv:astro-ph/0012055 [astro-ph]}
  \BibitemShut {NoStop}%
\bibitem [{\citenamefont {{Ruel}}\ \emph {et~al.}(2014)\citenamefont {{Ruel}},
  \citenamefont {{Bazin}}, \citenamefont {{Bayliss}}, \citenamefont
  {{Brodwin}}, \citenamefont {{Foley}}, \citenamefont {{Stalder}},
  \citenamefont {{Aird}}, \citenamefont {{Armstrong}}, \citenamefont {{Ashby}},
  \citenamefont {{Bautz}}, \citenamefont {{Benson}}, \citenamefont {{Bleem}},
  \citenamefont {{Bocquet}}, \citenamefont {{Carlstrom}}, \citenamefont
  {{Chang}}, \citenamefont {{Chapman}}, \citenamefont {{Cho}}, \citenamefont
  {{Clocchiatti}}, \citenamefont {{Crawford}}, \citenamefont {{Crites}},
  \citenamefont {{de Haan}}, \citenamefont {{Desai}}, \citenamefont {{Dobbs}},
  \citenamefont {{Dudley}}, \citenamefont {{Forman}}, \citenamefont {{George}},
  \citenamefont {{Gladders}}, \citenamefont {{Gonzalez}}, \citenamefont
  {{Halverson}}, \citenamefont {{Harrington}}, \citenamefont {{High}},
  \citenamefont {{Holder}}, \citenamefont {{Holzapfel}}, \citenamefont
  {{Hrubes}}, \citenamefont {{Jones}}, \citenamefont {{Joy}}, \citenamefont
  {{Keisler}}, \citenamefont {{Knox}}, \citenamefont {{Lee}}, \citenamefont
  {{Leitch}}, \citenamefont {{Liu}}, \citenamefont {{Lueker}}, \citenamefont
  {{Luong-Van}}, \citenamefont {{Mantz}}, \citenamefont {{Marrone}},
  \citenamefont {{McDonald}}, \citenamefont {{McMahon}}, \citenamefont
  {{Mehl}}, \citenamefont {{Meyer}}, \citenamefont {{Mocanu}}, \citenamefont
  {{Mohr}}, \citenamefont {{Montroy}}, \citenamefont {{Murray}}, \citenamefont
  {{Natoli}}, \citenamefont {{Nurgaliev}}, \citenamefont {{Padin}},
  \citenamefont {{Plagge}}, \citenamefont {{Pryke}}, \citenamefont
  {{Reichardt}}, \citenamefont {{Rest}}, \citenamefont {{Ruhl}}, \citenamefont
  {{Saliwanchik}}, \citenamefont {{Saro}}, \citenamefont {{Sayre}},
  \citenamefont {{Schaffer}}, \citenamefont {{Shaw}}, \citenamefont
  {{Shirokoff}}, \citenamefont {{Song}}, \citenamefont {{{\v{S}}uhada}},
  \citenamefont {{Spieler}}, \citenamefont {{Stanford}}, \citenamefont
  {{Staniszewski}}, \citenamefont {{Starsk}}, \citenamefont {{Story}},
  \citenamefont {{Stubbs}}, \citenamefont {{van Engelen}}, \citenamefont
  {{Vanderlinde}}, \citenamefont {{Vieira}}, \citenamefont {{Vikhlinin}},
  \citenamefont {{Williamson}}, \citenamefont {{Zahn}},\ and\ \citenamefont
  {{Zenteno}}}]{ruel_bazin_2014}%
  \BibitemOpen
  \bibfield  {author} {\bibinfo {author} {\bibfnamefont {J.}~\bibnamefont
  {{Ruel}}}, \bibinfo {author} {\bibfnamefont {G.}~\bibnamefont {{Bazin}}},
  \bibinfo {author} {\bibfnamefont {M.}~\bibnamefont {{Bayliss}}}, \bibinfo
  {author} {\bibfnamefont {M.}~\bibnamefont {{Brodwin}}}, \bibinfo {author}
  {\bibfnamefont {R.~J.}\ \bibnamefont {{Foley}}}, \bibinfo {author}
  {\bibfnamefont {B.}~\bibnamefont {{Stalder}}}, \bibinfo {author}
  {\bibfnamefont {K.~A.}\ \bibnamefont {{Aird}}}, \bibinfo {author}
  {\bibfnamefont {R.}~\bibnamefont {{Armstrong}}}, \bibinfo {author}
  {\bibfnamefont {M.~L.~N.}\ \bibnamefont {{Ashby}}}, \bibinfo {author}
  {\bibfnamefont {M.}~\bibnamefont {{Bautz}}}, \bibinfo {author} {\bibfnamefont
  {B.~A.}\ \bibnamefont {{Benson}}}, \bibinfo {author} {\bibfnamefont {L.~E.}\
  \bibnamefont {{Bleem}}}, \bibinfo {author} {\bibfnamefont {S.}~\bibnamefont
  {{Bocquet}}}, \bibinfo {author} {\bibfnamefont {J.~E.}\ \bibnamefont
  {{Carlstrom}}}, \bibinfo {author} {\bibfnamefont {C.~L.}\ \bibnamefont
  {{Chang}}}, \bibinfo {author} {\bibfnamefont {S.~C.}\ \bibnamefont
  {{Chapman}}}, \bibinfo {author} {\bibfnamefont {H.~M.}\ \bibnamefont
  {{Cho}}}, \bibinfo {author} {\bibfnamefont {A.}~\bibnamefont
  {{Clocchiatti}}}, \bibinfo {author} {\bibfnamefont {T.~M.}\ \bibnamefont
  {{Crawford}}}, \bibinfo {author} {\bibfnamefont {A.~T.}\ \bibnamefont
  {{Crites}}}, \bibinfo {author} {\bibfnamefont {T.}~\bibnamefont {{de Haan}}},
  \bibinfo {author} {\bibfnamefont {S.}~\bibnamefont {{Desai}}}, \bibinfo
  {author} {\bibfnamefont {M.~A.}\ \bibnamefont {{Dobbs}}}, \bibinfo {author}
  {\bibfnamefont {J.~P.}\ \bibnamefont {{Dudley}}}, \bibinfo {author}
  {\bibfnamefont {W.~R.}\ \bibnamefont {{Forman}}}, \bibinfo {author}
  {\bibfnamefont {E.~M.}\ \bibnamefont {{George}}}, \bibinfo {author}
  {\bibfnamefont {M.~D.}\ \bibnamefont {{Gladders}}}, \bibinfo {author}
  {\bibfnamefont {A.~H.}\ \bibnamefont {{Gonzalez}}}, \bibinfo {author}
  {\bibfnamefont {N.~W.}\ \bibnamefont {{Halverson}}}, \bibinfo {author}
  {\bibfnamefont {N.~L.}\ \bibnamefont {{Harrington}}}, \bibinfo {author}
  {\bibfnamefont {F.~W.}\ \bibnamefont {{High}}}, \bibinfo {author}
  {\bibfnamefont {G.~P.}\ \bibnamefont {{Holder}}}, \bibinfo {author}
  {\bibfnamefont {W.~L.}\ \bibnamefont {{Holzapfel}}}, \bibinfo {author}
  {\bibfnamefont {J.~D.}\ \bibnamefont {{Hrubes}}}, \bibinfo {author}
  {\bibfnamefont {C.}~\bibnamefont {{Jones}}}, \bibinfo {author} {\bibfnamefont
  {M.}~\bibnamefont {{Joy}}}, \bibinfo {author} {\bibfnamefont
  {R.}~\bibnamefont {{Keisler}}}, \bibinfo {author} {\bibfnamefont
  {L.}~\bibnamefont {{Knox}}}, \bibinfo {author} {\bibfnamefont {A.~T.}\
  \bibnamefont {{Lee}}}, \bibinfo {author} {\bibfnamefont {E.~M.}\ \bibnamefont
  {{Leitch}}}, \bibinfo {author} {\bibfnamefont {J.}~\bibnamefont {{Liu}}},
  \bibinfo {author} {\bibfnamefont {M.}~\bibnamefont {{Lueker}}}, \bibinfo
  {author} {\bibfnamefont {D.}~\bibnamefont {{Luong-Van}}}, \bibinfo {author}
  {\bibfnamefont {A.}~\bibnamefont {{Mantz}}}, \bibinfo {author} {\bibfnamefont
  {D.~P.}\ \bibnamefont {{Marrone}}}, \bibinfo {author} {\bibfnamefont
  {M.}~\bibnamefont {{McDonald}}}, \bibinfo {author} {\bibfnamefont {J.~J.}\
  \bibnamefont {{McMahon}}}, \bibinfo {author} {\bibfnamefont {J.}~\bibnamefont
  {{Mehl}}}, \bibinfo {author} {\bibfnamefont {S.~S.}\ \bibnamefont {{Meyer}}},
  \bibinfo {author} {\bibfnamefont {L.}~\bibnamefont {{Mocanu}}}, \bibinfo
  {author} {\bibfnamefont {J.~J.}\ \bibnamefont {{Mohr}}}, \bibinfo {author}
  {\bibfnamefont {T.~E.}\ \bibnamefont {{Montroy}}}, \bibinfo {author}
  {\bibfnamefont {S.~S.}\ \bibnamefont {{Murray}}}, \bibinfo {author}
  {\bibfnamefont {T.}~\bibnamefont {{Natoli}}}, \bibinfo {author}
  {\bibfnamefont {D.}~\bibnamefont {{Nurgaliev}}}, \bibinfo {author}
  {\bibfnamefont {S.}~\bibnamefont {{Padin}}}, \bibinfo {author} {\bibfnamefont
  {T.}~\bibnamefont {{Plagge}}}, \bibinfo {author} {\bibfnamefont
  {C.}~\bibnamefont {{Pryke}}}, \bibinfo {author} {\bibfnamefont {C.~L.}\
  \bibnamefont {{Reichardt}}}, \bibinfo {author} {\bibfnamefont
  {A.}~\bibnamefont {{Rest}}}, \bibinfo {author} {\bibfnamefont {J.~E.}\
  \bibnamefont {{Ruhl}}}, \bibinfo {author} {\bibfnamefont {B.~R.}\
  \bibnamefont {{Saliwanchik}}}, \bibinfo {author} {\bibfnamefont
  {A.}~\bibnamefont {{Saro}}}, \bibinfo {author} {\bibfnamefont {J.~T.}\
  \bibnamefont {{Sayre}}}, \bibinfo {author} {\bibfnamefont {K.~K.}\
  \bibnamefont {{Schaffer}}}, \bibinfo {author} {\bibfnamefont
  {L.}~\bibnamefont {{Shaw}}}, \bibinfo {author} {\bibfnamefont
  {E.}~\bibnamefont {{Shirokoff}}}, \bibinfo {author} {\bibfnamefont
  {J.}~\bibnamefont {{Song}}}, \bibinfo {author} {\bibfnamefont
  {R.}~\bibnamefont {{{\v{S}}uhada}}}, \bibinfo {author} {\bibfnamefont
  {H.~G.}\ \bibnamefont {{Spieler}}}, \bibinfo {author} {\bibfnamefont {S.~A.}\
  \bibnamefont {{Stanford}}}, \bibinfo {author} {\bibfnamefont
  {Z.}~\bibnamefont {{Staniszewski}}}, \bibinfo {author} {\bibfnamefont
  {A.~A.}\ \bibnamefont {{Starsk}}}, \bibinfo {author} {\bibfnamefont
  {K.}~\bibnamefont {{Story}}}, \bibinfo {author} {\bibfnamefont {C.~W.}\
  \bibnamefont {{Stubbs}}}, \bibinfo {author} {\bibfnamefont {A.}~\bibnamefont
  {{van Engelen}}}, \bibinfo {author} {\bibfnamefont {K.}~\bibnamefont
  {{Vanderlinde}}}, \bibinfo {author} {\bibfnamefont {J.~D.}\ \bibnamefont
  {{Vieira}}}, \bibinfo {author} {\bibfnamefont {A.}~\bibnamefont
  {{Vikhlinin}}}, \bibinfo {author} {\bibfnamefont {R.}~\bibnamefont
  {{Williamson}}}, \bibinfo {author} {\bibfnamefont {O.}~\bibnamefont
  {{Zahn}}},\ and\ \bibinfo {author} {\bibfnamefont {A.}~\bibnamefont
  {{Zenteno}}},\ }\href {https://doi.org/10.1088/0004-637X/792/1/45} {\bibfield
   {journal} {\bibinfo  {journal} {\apj}\ }\textbf {\bibinfo {volume} {792}},\
  \bibinfo {eid} {45} (\bibinfo {year} {2014})},\ \Eprint
  {https://arxiv.org/abs/1311.4953} {arXiv:1311.4953 [astro-ph.CO]}
  \BibitemShut {NoStop}%
\bibitem [{\citenamefont {{Wen}}(2003)}]{wen_2003}%
  \BibitemOpen
  \bibfield  {author} {\bibinfo {author} {\bibfnamefont {L.}~\bibnamefont
  {{Wen}}},\ }\href {https://doi.org/10.1086/378794} {\bibfield  {journal}
  {\bibinfo  {journal} {\apj}\ }\textbf {\bibinfo {volume} {598}},\ \bibinfo
  {pages} {419} (\bibinfo {year} {2003})},\ \Eprint
  {https://arxiv.org/abs/astro-ph/0211492} {arXiv:astro-ph/0211492 [astro-ph]}
  \BibitemShut {NoStop}%
\bibitem [{\citenamefont {{Naoz}}(2016)}]{naoz_2016}%
  \BibitemOpen
  \bibfield  {author} {\bibinfo {author} {\bibfnamefont {S.}~\bibnamefont
  {{Naoz}}},\ }\href {https://doi.org/10.1146/annurev-astro-081915-023315}
  {\bibfield  {journal} {\bibinfo  {journal} {\araa}\ }\textbf {\bibinfo
  {volume} {54}},\ \bibinfo {pages} {441} (\bibinfo {year} {2016})},\ \Eprint
  {https://arxiv.org/abs/1601.07175} {arXiv:1601.07175 [astro-ph.EP]}
  \BibitemShut {NoStop}%
\bibitem [{\citenamefont {{Meiron}}\ \emph {et~al.}(2017)\citenamefont
  {{Meiron}}, \citenamefont {{Kocsis}},\ and\ \citenamefont
  {{Loeb}}}]{meiron_kocsis_2017}%
  \BibitemOpen
  \bibfield  {author} {\bibinfo {author} {\bibfnamefont {Y.}~\bibnamefont
  {{Meiron}}}, \bibinfo {author} {\bibfnamefont {B.}~\bibnamefont {{Kocsis}}},\
  and\ \bibinfo {author} {\bibfnamefont {A.}~\bibnamefont {{Loeb}}},\ }\href
  {https://doi.org/10.3847/1538-4357/834/2/200} {\bibfield  {journal} {\bibinfo
   {journal} {\apj}\ }\textbf {\bibinfo {volume} {834}},\ \bibinfo {eid} {200}
  (\bibinfo {year} {2017})},\ \Eprint {https://arxiv.org/abs/1604.02148}
  {arXiv:1604.02148 [astro-ph.HE]} \BibitemShut {NoStop}%
\bibitem [{\citenamefont {{Arca Sedda}}(2020)}]{arca-sedda_2020}%
  \BibitemOpen
  \bibfield  {author} {\bibinfo {author} {\bibfnamefont {M.}~\bibnamefont
  {{Arca Sedda}}},\ }\href {https://doi.org/10.3847/1538-4357/ab723b}
  {\bibfield  {journal} {\bibinfo  {journal} {\apj}\ }\textbf {\bibinfo
  {volume} {891}},\ \bibinfo {eid} {47} (\bibinfo {year} {2020})},\ \Eprint
  {https://arxiv.org/abs/2002.04037} {arXiv:2002.04037 [astro-ph.GA]}
  \BibitemShut {NoStop}%
\bibitem [{\citenamefont {{Antonini}}\ and\ \citenamefont
  {{Perets}}(2012)}]{antonini_perets_2012}%
  \BibitemOpen
  \bibfield  {author} {\bibinfo {author} {\bibfnamefont {F.}~\bibnamefont
  {{Antonini}}}\ and\ \bibinfo {author} {\bibfnamefont {H.~B.}\ \bibnamefont
  {{Perets}}},\ }\href {https://doi.org/10.1088/0004-637X/757/1/27} {\bibfield
  {journal} {\bibinfo  {journal} {\apj}\ }\textbf {\bibinfo {volume} {757}},\
  \bibinfo {eid} {27} (\bibinfo {year} {2012})},\ \Eprint
  {https://arxiv.org/abs/1203.2938} {arXiv:1203.2938} \BibitemShut {NoStop}%
\bibitem [{\citenamefont {{McKernan}}\ \emph {et~al.}(2012)\citenamefont
  {{McKernan}}, \citenamefont {{Ford}}, \citenamefont {{Lyra}},\ and\
  \citenamefont {{Perets}}}]{mckernan_ford_2012}%
  \BibitemOpen
  \bibfield  {author} {\bibinfo {author} {\bibfnamefont {B.}~\bibnamefont
  {{McKernan}}}, \bibinfo {author} {\bibfnamefont {K.~E.~S.}\ \bibnamefont
  {{Ford}}}, \bibinfo {author} {\bibfnamefont {W.}~\bibnamefont {{Lyra}}},\
  and\ \bibinfo {author} {\bibfnamefont {H.~B.}\ \bibnamefont {{Perets}}},\
  }\href {https://doi.org/10.1111/j.1365-2966.2012.21486.x} {\bibfield
  {journal} {\bibinfo  {journal} {\mnras}\ }\textbf {\bibinfo {volume} {425}},\
  \bibinfo {pages} {460} (\bibinfo {year} {2012})},\ \Eprint
  {https://arxiv.org/abs/1206.2309} {arXiv:1206.2309 [astro-ph.GA]}
  \BibitemShut {NoStop}%
\bibitem [{\citenamefont {{Addison}}\ \emph {et~al.}(2019)\citenamefont
  {{Addison}}, \citenamefont {{Gracia-Linares}}, \citenamefont {{Laguna}},\
  and\ \citenamefont {{Larson}}}]{addison_gracia-linares_2019}%
  \BibitemOpen
  \bibfield  {author} {\bibinfo {author} {\bibfnamefont {E.}~\bibnamefont
  {{Addison}}}, \bibinfo {author} {\bibfnamefont {M.}~\bibnamefont
  {{Gracia-Linares}}}, \bibinfo {author} {\bibfnamefont {P.}~\bibnamefont
  {{Laguna}}},\ and\ \bibinfo {author} {\bibfnamefont {S.~L.}\ \bibnamefont
  {{Larson}}},\ }\href {https://doi.org/10.1007/s10714-019-2523-4} {\bibfield
  {journal} {\bibinfo  {journal} {General Relativity and Gravitation}\ }\textbf
  {\bibinfo {volume} {51}},\ \bibinfo {eid} {38} (\bibinfo {year}
  {2019})}\BibitemShut {NoStop}%
\bibitem [{\citenamefont {{Bartos}}\ \emph {et~al.}(2017)\citenamefont
  {{Bartos}}, \citenamefont {{Kocsis}}, \citenamefont {{Haiman}},\ and\
  \citenamefont {{M{\'a}rka}}}]{bartos_kocsis_2017}%
  \BibitemOpen
  \bibfield  {author} {\bibinfo {author} {\bibfnamefont {I.}~\bibnamefont
  {{Bartos}}}, \bibinfo {author} {\bibfnamefont {B.}~\bibnamefont {{Kocsis}}},
  \bibinfo {author} {\bibfnamefont {Z.}~\bibnamefont {{Haiman}}},\ and\
  \bibinfo {author} {\bibfnamefont {S.}~\bibnamefont {{M{\'a}rka}}},\ }\href
  {https://doi.org/10.3847/1538-4357/835/2/165} {\bibfield  {journal} {\bibinfo
   {journal} {\apj}\ }\textbf {\bibinfo {volume} {835}},\ \bibinfo {eid} {165}
  (\bibinfo {year} {2017})},\ \Eprint {https://arxiv.org/abs/1602.03831}
  {arXiv:1602.03831 [astro-ph.HE]} \BibitemShut {NoStop}%
\bibitem [{\citenamefont {{Stone}}\ \emph {et~al.}(2017)\citenamefont
  {{Stone}}, \citenamefont {{Metzger}},\ and\ \citenamefont
  {{Haiman}}}]{stone_metzger_2017}%
  \BibitemOpen
  \bibfield  {author} {\bibinfo {author} {\bibfnamefont {N.~C.}\ \bibnamefont
  {{Stone}}}, \bibinfo {author} {\bibfnamefont {B.~D.}\ \bibnamefont
  {{Metzger}}},\ and\ \bibinfo {author} {\bibfnamefont {Z.}~\bibnamefont
  {{Haiman}}},\ }\href {https://doi.org/10.1093/mnras/stw2260} {\bibfield
  {journal} {\bibinfo  {journal} {\mnras}\ }\textbf {\bibinfo {volume} {464}},\
  \bibinfo {pages} {946} (\bibinfo {year} {2017})},\ \Eprint
  {https://arxiv.org/abs/1602.04226} {arXiv:1602.04226} \BibitemShut {NoStop}%
\bibitem [{\citenamefont {{Tagawa}}\ \emph {et~al.}(2020)\citenamefont
  {{Tagawa}}, \citenamefont {{Haiman}},\ and\ \citenamefont
  {{Kocsis}}}]{tagawa_haiman_2020}%
  \BibitemOpen
  \bibfield  {author} {\bibinfo {author} {\bibfnamefont {H.}~\bibnamefont
  {{Tagawa}}}, \bibinfo {author} {\bibfnamefont {Z.}~\bibnamefont {{Haiman}}},\
  and\ \bibinfo {author} {\bibfnamefont {B.}~\bibnamefont {{Kocsis}}},\ }\href
  {https://doi.org/10.3847/1538-4357/ab9b8c} {\bibfield  {journal} {\bibinfo
  {journal} {\apj}\ }\textbf {\bibinfo {volume} {898}},\ \bibinfo {eid} {25}
  (\bibinfo {year} {2020})},\ \Eprint {https://arxiv.org/abs/1912.08218}
  {arXiv:1912.08218 [astro-ph.GA]} \BibitemShut {NoStop}%
\bibitem [{\citenamefont {{Chen}}\ \emph {et~al.}(2019)\citenamefont {{Chen}},
  \citenamefont {{Li}},\ and\ \citenamefont {{Cao}}}]{chen_li_2017}%
  \BibitemOpen
  \bibfield  {author} {\bibinfo {author} {\bibfnamefont {X.}~\bibnamefont
  {{Chen}}}, \bibinfo {author} {\bibfnamefont {S.}~\bibnamefont {{Li}}},\ and\
  \bibinfo {author} {\bibfnamefont {Z.}~\bibnamefont {{Cao}}},\ }\href
  {https://doi.org/10.1093/mnrasl/slz046} {\bibfield  {journal} {\bibinfo
  {journal} {Monthly Notices of the Royal Astronomical Society: Letters}\
  }\textbf {\bibinfo {volume} {485}},\ \bibinfo {pages} {L141} (\bibinfo {year}
  {2019})},\ \Eprint {https://arxiv.org/abs/1703.10543} {arXiv:1703.10543
  [astro-ph.HE]} \BibitemShut {NoStop}%
\bibitem [{\citenamefont {{Chen}}\ and\ \citenamefont
  {{Han}}(2018)}]{chen_han_2018}%
  \BibitemOpen
  \bibfield  {author} {\bibinfo {author} {\bibfnamefont {X.}~\bibnamefont
  {{Chen}}}\ and\ \bibinfo {author} {\bibfnamefont {W.-B.}\ \bibnamefont
  {{Han}}},\ }\href {https://doi.org/10.1038/s42005-018-0053-0} {\bibfield
  {journal} {\bibinfo  {journal} {Communications Physics}\ }\textbf {\bibinfo
  {volume} {1}},\ \bibinfo {eid} {53} (\bibinfo {year} {2018})},\ \Eprint
  {https://arxiv.org/abs/1801.05780} {arXiv:1801.05780 [astro-ph.HE]}
  \BibitemShut {NoStop}%
\bibitem [{\citenamefont {{Han}}\ and\ \citenamefont
  {{Chen}}(2019)}]{han_chen_2018}%
  \BibitemOpen
  \bibfield  {author} {\bibinfo {author} {\bibfnamefont {W.-B.}\ \bibnamefont
  {{Han}}}\ and\ \bibinfo {author} {\bibfnamefont {X.}~\bibnamefont {{Chen}}},\
  }\href {https://doi.org/10.1093/mnrasl/slz021} {\bibfield  {journal}
  {\bibinfo  {journal} {\mnras}\ }\textbf {\bibinfo {volume} {485}},\ \bibinfo
  {pages} {L29} (\bibinfo {year} {2019})},\ \Eprint
  {https://arxiv.org/abs/1801.07060} {arXiv:1801.07060 [gr-qc]} \BibitemShut
  {NoStop}%
\bibitem [{\citenamefont {{Ruiz}}\ \emph {et~al.}(2008)\citenamefont {{Ruiz}},
  \citenamefont {{Alcubierre}}, \citenamefont {{N{\'u}{\~n}ez}},\ and\
  \citenamefont {{Takahashi}}}]{ruiz_alcubierre_2008}%
  \BibitemOpen
  \bibfield  {author} {\bibinfo {author} {\bibfnamefont {M.}~\bibnamefont
  {{Ruiz}}}, \bibinfo {author} {\bibfnamefont {M.}~\bibnamefont
  {{Alcubierre}}}, \bibinfo {author} {\bibfnamefont {D.}~\bibnamefont
  {{N{\'u}{\~n}ez}}},\ and\ \bibinfo {author} {\bibfnamefont {R.}~\bibnamefont
  {{Takahashi}}},\ }\href {https://doi.org/10.1007/s10714-008-0684-7}
  {\bibfield  {journal} {\bibinfo  {journal} {General Relativity and
  Gravitation}\ }\textbf {\bibinfo {volume} {40}},\ \bibinfo {pages} {2467}
  (\bibinfo {year} {2008})}\BibitemShut {NoStop}%
\bibitem [{\citenamefont {{Goldberg}}\ \emph {et~al.}(1967)\citenamefont
  {{Goldberg}}, \citenamefont {{Macfarlane}}, \citenamefont {{Newman}},
  \citenamefont {{Rohrlich}},\ and\ \citenamefont
  {{Sudarshan}}}]{goldberg_macfarlane_1967}%
  \BibitemOpen
  \bibfield  {author} {\bibinfo {author} {\bibfnamefont {J.~N.}\ \bibnamefont
  {{Goldberg}}}, \bibinfo {author} {\bibfnamefont {A.~J.}\ \bibnamefont
  {{Macfarlane}}}, \bibinfo {author} {\bibfnamefont {E.~T.}\ \bibnamefont
  {{Newman}}}, \bibinfo {author} {\bibfnamefont {F.}~\bibnamefont
  {{Rohrlich}}},\ and\ \bibinfo {author} {\bibfnamefont {E.~C.~G.}\
  \bibnamefont {{Sudarshan}}},\ }\href {https://doi.org/10.1063/1.1705135}
  {\bibfield  {journal} {\bibinfo  {journal} {Journal of Mathematical Physics}\
  }\textbf {\bibinfo {volume} {8}},\ \bibinfo {pages} {2155} (\bibinfo {year}
  {1967})}\BibitemShut {NoStop}%
\bibitem [{\citenamefont {Thorne}(1980)}]{thorne_1980}%
  \BibitemOpen
  \bibfield  {author} {\bibinfo {author} {\bibfnamefont {K.~S.}\ \bibnamefont
  {Thorne}},\ }\href {https://doi.org/10.1103/RevModPhys.52.299} {\bibfield
  {journal} {\bibinfo  {journal} {Rev. Mod. Phys.}\ }\textbf {\bibinfo {volume}
  {52}},\ \bibinfo {pages} {299} (\bibinfo {year} {1980})}\BibitemShut
  {NoStop}%
\bibitem [{\citenamefont {{Torres-Orjuela}}\ \emph {et~al.}(2019)\citenamefont
  {{Torres-Orjuela}}, \citenamefont {{Chen}}, \citenamefont {{Cao}},
  \citenamefont {{Amaro-Seoane}},\ and\ \citenamefont
  {{Peng}}}]{torres-orjuela_chen_2019}%
  \BibitemOpen
  \bibfield  {author} {\bibinfo {author} {\bibfnamefont {A.}~\bibnamefont
  {{Torres-Orjuela}}}, \bibinfo {author} {\bibfnamefont {X.}~\bibnamefont
  {{Chen}}}, \bibinfo {author} {\bibfnamefont {Z.}~\bibnamefont {{Cao}}},
  \bibinfo {author} {\bibfnamefont {P.}~\bibnamefont {{Amaro-Seoane}}},\ and\
  \bibinfo {author} {\bibfnamefont {P.}~\bibnamefont {{Peng}}},\ }\href
  {https://doi.org/10.1103/PhysRevD.100.063012} {\bibfield  {journal} {\bibinfo
   {journal} {\prd}\ }\textbf {\bibinfo {volume} {100}},\ \bibinfo {eid}
  {063012} (\bibinfo {year} {2019})},\ \Eprint
  {https://arxiv.org/abs/1806.09857} {arXiv:1806.09857 [astro-ph.HE]}
  \BibitemShut {NoStop}%
\bibitem [{\citenamefont {{Gualtieri}}\ \emph {et~al.}(2008)\citenamefont
  {{Gualtieri}}, \citenamefont {{Berti}}, \citenamefont {{Cardoso}},\ and\
  \citenamefont {{Sperhake}}}]{gualtieri_berti_2008}%
  \BibitemOpen
  \bibfield  {author} {\bibinfo {author} {\bibfnamefont {L.}~\bibnamefont
  {{Gualtieri}}}, \bibinfo {author} {\bibfnamefont {E.}~\bibnamefont
  {{Berti}}}, \bibinfo {author} {\bibfnamefont {V.}~\bibnamefont {{Cardoso}}},\
  and\ \bibinfo {author} {\bibfnamefont {U.}~\bibnamefont {{Sperhake}}},\
  }\href {https://doi.org/10.1103/PhysRevD.78.044024} {\bibfield  {journal}
  {\bibinfo  {journal} {\prd}\ }\textbf {\bibinfo {volume} {78}},\ \bibinfo
  {eid} {044024} (\bibinfo {year} {2008})},\ \Eprint
  {https://arxiv.org/abs/0805.1017} {arXiv:0805.1017 [gr-qc]} \BibitemShut
  {NoStop}%
\bibitem [{\citenamefont {{Boyle}}(2016)}]{boyle_2016}%
  \BibitemOpen
  \bibfield  {author} {\bibinfo {author} {\bibfnamefont {M.}~\bibnamefont
  {{Boyle}}},\ }\href {https://doi.org/10.1103/PhysRevD.93.084031} {\bibfield
  {journal} {\bibinfo  {journal} {\prd}\ }\textbf {\bibinfo {volume} {93}},\
  \bibinfo {eid} {084031} (\bibinfo {year} {2016})},\ \Eprint
  {https://arxiv.org/abs/1509.00862} {arXiv:1509.00862 [gr-qc]} \BibitemShut
  {NoStop}%
\bibitem [{\citenamefont {{Woodford}}\ \emph {et~al.}(2019)\citenamefont
  {{Woodford}}, \citenamefont {{Boyle}},\ and\ \citenamefont
  {{Pfeiffer}}}]{woodford_boyle_2019}%
  \BibitemOpen
  \bibfield  {author} {\bibinfo {author} {\bibfnamefont {C.~J.}\ \bibnamefont
  {{Woodford}}}, \bibinfo {author} {\bibfnamefont {M.}~\bibnamefont
  {{Boyle}}},\ and\ \bibinfo {author} {\bibfnamefont {H.~P.}\ \bibnamefont
  {{Pfeiffer}}},\ }\href {https://doi.org/10.1103/PhysRevD.100.124010}
  {\bibfield  {journal} {\bibinfo  {journal} {\prd}\ }\textbf {\bibinfo
  {volume} {100}},\ \bibinfo {eid} {124010} (\bibinfo {year} {2019})},\ \Eprint
  {https://arxiv.org/abs/1904.04842} {arXiv:1904.04842 [gr-qc]} \BibitemShut
  {NoStop}%
\bibitem [{\citenamefont {{Misner}}\ \emph {et~al.}(2017)\citenamefont
  {{Misner}}, \citenamefont {{Thorne}},\ and\ \citenamefont
  {{Wheeler}}}]{misner_thorne_1973}%
  \BibitemOpen
  \bibfield  {author} {\bibinfo {author} {\bibfnamefont {C.~W.}\ \bibnamefont
  {{Misner}}}, \bibinfo {author} {\bibfnamefont {K.~S.}\ \bibnamefont
  {{Thorne}}},\ and\ \bibinfo {author} {\bibfnamefont {J.~A.}\ \bibnamefont
  {{Wheeler}}},\ }\href@noop {} {\emph {\bibinfo {title} {Gravitation, by
  Charles W.~Misner, Kip S.~Thorne, and John Archibald Wheeler.~ISBN:
  978-0-691-17779-3.~Princeton NJ: Princeton University Press, 2017.}}}\
  (\bibinfo  {publisher} {Princeton University Press},\ \bibinfo {address} {San
  Francisco, {CA}},\ \bibinfo {year} {2017})\BibitemShut {NoStop}%
\bibitem [{\citenamefont {{Torres-Orjuela}}\ \emph {et~al.}(2020)\citenamefont
  {{Torres-Orjuela}}, \citenamefont {{Chen}},\ and\ \citenamefont
  {{Amaro-Seoane}}}]{torres-orjuela_chen_2020a}%
  \BibitemOpen
  \bibfield  {author} {\bibinfo {author} {\bibfnamefont {A.}~\bibnamefont
  {{Torres-Orjuela}}}, \bibinfo {author} {\bibfnamefont {X.}~\bibnamefont
  {{Chen}}},\ and\ \bibinfo {author} {\bibfnamefont {P.}~\bibnamefont
  {{Amaro-Seoane}}},\ }\href {https://doi.org/10.1103/PhysRevD.101.083028}
  {\bibfield  {journal} {\bibinfo  {journal} {\prd}\ }\textbf {\bibinfo
  {volume} {101}},\ \bibinfo {eid} {083028} (\bibinfo {year} {2020})},\ \Eprint
  {https://arxiv.org/abs/2001.00721} {arXiv:2001.00721 [astro-ph.HE]}
  \BibitemShut {NoStop}%
\bibitem [{\citenamefont {Jackson}(1999)}]{jackson_2009}%
  \BibitemOpen
  \bibfield  {author} {\bibinfo {author} {\bibfnamefont {J.~D.}\ \bibnamefont
  {Jackson}},\ }\href {http://cdsweb.cern.ch/record/490457} {\emph {\bibinfo
  {title} {Classical electrodynamics}}},\ \bibinfo {edition} {3rd}\ ed.\
  (\bibinfo  {publisher} {Wiley},\ \bibinfo {address} {New York, {NY}},\
  \bibinfo {year} {1999})\BibitemShut {NoStop}%
\bibitem [{\citenamefont {{Thorne}}(1987)}]{thorne_1987}%
  \BibitemOpen
  \bibfield  {author} {\bibinfo {author} {\bibfnamefont {K.~S.}\ \bibnamefont
  {{Thorne}}},\ }\href {https://doi.org/10.1126/science.240.4855.1069} {\emph
  {\bibinfo {title} {Three Hundred Years of Gravitation}}},\ edited by\
  \bibinfo {editor} {\bibfnamefont {S.~W.}\ \bibnamefont {{Hawking}}}\ and\
  \bibinfo {editor} {\bibfnamefont {W.}~\bibnamefont {{Israel}}}\ (\bibinfo
  {publisher} {Cambridge University Press},\ \bibinfo {address} {New York,
  {NY}},\ \bibinfo {year} {1987})\ pp.\ \bibinfo {pages} {330--458}\BibitemShut
  {NoStop}%
\bibitem [{\citenamefont {{The LIGO Scientific Collaboration}}\ and\
  \citenamefont {{the Virgo Collaboration}}(2019)}]{GWTC1}%
  \BibitemOpen
  \bibfield  {author} {\bibinfo {author} {\bibnamefont {{The LIGO Scientific
  Collaboration}}}\ and\ \bibinfo {author} {\bibnamefont {{the Virgo
  Collaboration}}},\ }\href {https://doi.org/10.1103/PhysRevX.9.031040}
  {\bibfield  {journal} {\bibinfo  {journal} {Physical Review X}\ }\textbf
  {\bibinfo {volume} {9}},\ \bibinfo {eid} {031040} (\bibinfo {year} {2019})},\
  \Eprint {https://arxiv.org/abs/1811.12907} {arXiv:1811.12907 [astro-ph.HE]}
  \BibitemShut {NoStop}%
\bibitem [{\citenamefont {{The LIGO Scientific Collaboration}}\ and\
  \citenamefont {{the Virgo Collaboration}}(2021)}]{GWTC2}%
  \BibitemOpen
  \bibfield  {author} {\bibinfo {author} {\bibnamefont {{The LIGO Scientific
  Collaboration}}}\ and\ \bibinfo {author} {\bibnamefont {{the Virgo
  Collaboration}}},\ }\href {https://doi.org/10.1103/PhysRevX.11.021053}
  {\bibfield  {journal} {\bibinfo  {journal} {Physical Review X}\ }\textbf
  {\bibinfo {volume} {11}},\ \bibinfo {eid} {021053} (\bibinfo {year}
  {2021})},\ \Eprint {https://arxiv.org/abs/2010.14527} {arXiv:2010.14527
  [gr-qc]} \BibitemShut {NoStop}%
\bibitem [{\citenamefont {{Schutz}}(1986)}]{schutz_1986}%
  \BibitemOpen
  \bibfield  {author} {\bibinfo {author} {\bibfnamefont {B.~F.}\ \bibnamefont
  {{Schutz}}},\ }\href {https://doi.org/10.1038/323310a0} {\bibfield  {journal}
  {\bibinfo  {journal} {\nat}\ }\textbf {\bibinfo {volume} {323}},\ \bibinfo
  {pages} {310} (\bibinfo {year} {1986})}\BibitemShut {NoStop}%
\bibitem [{\citenamefont {{Cutler}}\ and\ \citenamefont
  {{Flanagan}}(1994)}]{cutler_flanagan_1994}%
  \BibitemOpen
  \bibfield  {author} {\bibinfo {author} {\bibfnamefont {C.}~\bibnamefont
  {{Cutler}}}\ and\ \bibinfo {author} {\bibfnamefont {{\'E}.~E.}\ \bibnamefont
  {{Flanagan}}},\ }\href {https://doi.org/10.1103/PhysRevD.49.2658} {\bibfield
  {journal} {\bibinfo  {journal} {\prd}\ }\textbf {\bibinfo {volume} {49}},\
  \bibinfo {pages} {2658} (\bibinfo {year} {1994})},\ \Eprint
  {https://arxiv.org/abs/gr-qc/9402014} {arXiv:gr-qc/9402014 [gr-qc]}
  \BibitemShut {NoStop}%
\bibitem [{\citenamefont {{Torres-Orjuela}}\ \emph {et~al.}(2021)\citenamefont
  {{Torres-Orjuela}}, \citenamefont {{Amaro Seoane}}, \citenamefont {{Xuan}},
  \citenamefont {{Chua}}, \citenamefont {{Rosell}},\ and\ \citenamefont
  {{Chen}}}]{torres-orjuela_amaro-seoane_2020}%
  \BibitemOpen
  \bibfield  {author} {\bibinfo {author} {\bibfnamefont {A.}~\bibnamefont
  {{Torres-Orjuela}}}, \bibinfo {author} {\bibfnamefont {P.}~\bibnamefont
  {{Amaro Seoane}}}, \bibinfo {author} {\bibfnamefont {Z.}~\bibnamefont
  {{Xuan}}}, \bibinfo {author} {\bibfnamefont {A.~J.~K.}\ \bibnamefont
  {{Chua}}}, \bibinfo {author} {\bibfnamefont {M.~J.~B.}\ \bibnamefont
  {{Rosell}}},\ and\ \bibinfo {author} {\bibfnamefont {X.}~\bibnamefont
  {{Chen}}},\ }\href {https://doi.org/10.1103/PhysRevLett.127.041102}
  {\bibfield  {journal} {\bibinfo  {journal} {\prl}\ }\textbf {\bibinfo
  {volume} {127}},\ \bibinfo {eid} {041102} (\bibinfo {year} {2021})},\ \Eprint
  {https://arxiv.org/abs/2010.15842} {arXiv:2010.15842 [gr-qc]} \BibitemShut
  {NoStop}%
\bibitem [{\citenamefont {{Lindblom}}\ \emph {et~al.}(2008)\citenamefont
  {{Lindblom}}, \citenamefont {{Owen}},\ and\ \citenamefont
  {{Brown}}}]{lindblom_owen_2008}%
  \BibitemOpen
  \bibfield  {author} {\bibinfo {author} {\bibfnamefont {L.}~\bibnamefont
  {{Lindblom}}}, \bibinfo {author} {\bibfnamefont {B.~J.}\ \bibnamefont
  {{Owen}}},\ and\ \bibinfo {author} {\bibfnamefont {D.~A.}\ \bibnamefont
  {{Brown}}},\ }\href {https://doi.org/10.1103/PhysRevD.78.124020} {\bibfield
  {journal} {\bibinfo  {journal} {\prd}\ }\textbf {\bibinfo {volume} {78}},\
  \bibinfo {eid} {124020} (\bibinfo {year} {2008})},\ \Eprint
  {https://arxiv.org/abs/0809.3844} {arXiv:0809.3844 [gr-qc]} \BibitemShut
  {NoStop}%
\bibitem [{\citenamefont {{Arun}}\ \emph {et~al.}(2009)\citenamefont {{Arun}},
  \citenamefont {{Buonanno}}, \citenamefont {{Faye}},\ and\ \citenamefont
  {{Ochsner}}}]{arun_buonanno_2009}%
  \BibitemOpen
  \bibfield  {author} {\bibinfo {author} {\bibfnamefont {K.~G.}\ \bibnamefont
  {{Arun}}}, \bibinfo {author} {\bibfnamefont {A.}~\bibnamefont {{Buonanno}}},
  \bibinfo {author} {\bibfnamefont {G.}~\bibnamefont {{Faye}}},\ and\ \bibinfo
  {author} {\bibfnamefont {E.}~\bibnamefont {{Ochsner}}},\ }\href
  {https://doi.org/10.1103/PhysRevD.79.104023} {\bibfield  {journal} {\bibinfo
  {journal} {\prd}\ }\textbf {\bibinfo {volume} {79}},\ \bibinfo {eid} {104023}
  (\bibinfo {year} {2009})},\ \Eprint {https://arxiv.org/abs/0810.5336}
  {arXiv:0810.5336 [gr-qc]} \BibitemShut {NoStop}%
\bibitem [{\citenamefont {Nitz}\ \emph {et~al.}(2020)\citenamefont {Nitz},
  \citenamefont {Harry}, \citenamefont {Brown}, \citenamefont {Biwer},
  \citenamefont {Willis}, \citenamefont {Canton}, \citenamefont {Capano},
  \citenamefont {Pekowsky}, \citenamefont {Dent}, \citenamefont {Williamson},
  \citenamefont {Davies}, \citenamefont {De}, \citenamefont {Cabero},
  \citenamefont {Machenschalk}, \citenamefont {Kumar}, \citenamefont {Reyes},
  \citenamefont {Macleod}, \citenamefont {Pannarale}, \citenamefont {dfinstad},
  \citenamefont {Massinger}, \citenamefont {Tápai}, \citenamefont {Singer},
  \citenamefont {Khan}, \citenamefont {Fairhurst}, \citenamefont {Kumar},
  \citenamefont {Nielsen}, \citenamefont {SSingh087}, \citenamefont {shasvath},
  \citenamefont {Dorrington},\ and\ \citenamefont {Gadre}}]{pycbc_2020}%
  \BibitemOpen
  \bibfield  {author} {\bibinfo {author} {\bibfnamefont {A.}~\bibnamefont
  {Nitz}}, \bibinfo {author} {\bibfnamefont {I.}~\bibnamefont {Harry}},
  \bibinfo {author} {\bibfnamefont {D.}~\bibnamefont {Brown}}, \bibinfo
  {author} {\bibfnamefont {C.~M.}\ \bibnamefont {Biwer}}, \bibinfo {author}
  {\bibfnamefont {J.}~\bibnamefont {Willis}}, \bibinfo {author} {\bibfnamefont
  {T.~D.}\ \bibnamefont {Canton}}, \bibinfo {author} {\bibfnamefont
  {C.}~\bibnamefont {Capano}}, \bibinfo {author} {\bibfnamefont
  {L.}~\bibnamefont {Pekowsky}}, \bibinfo {author} {\bibfnamefont
  {T.}~\bibnamefont {Dent}}, \bibinfo {author} {\bibfnamefont {A.~R.}\
  \bibnamefont {Williamson}}, \bibinfo {author} {\bibfnamefont {G.~S.}\
  \bibnamefont {Davies}}, \bibinfo {author} {\bibfnamefont {S.}~\bibnamefont
  {De}}, \bibinfo {author} {\bibfnamefont {M.}~\bibnamefont {Cabero}}, \bibinfo
  {author} {\bibfnamefont {B.}~\bibnamefont {Machenschalk}}, \bibinfo {author}
  {\bibfnamefont {P.}~\bibnamefont {Kumar}}, \bibinfo {author} {\bibfnamefont
  {S.}~\bibnamefont {Reyes}}, \bibinfo {author} {\bibfnamefont
  {D.}~\bibnamefont {Macleod}}, \bibinfo {author} {\bibfnamefont
  {F.}~\bibnamefont {Pannarale}}, \bibinfo {author} {\bibnamefont {dfinstad}},
  \bibinfo {author} {\bibfnamefont {T.}~\bibnamefont {Massinger}}, \bibinfo
  {author} {\bibfnamefont {M.}~\bibnamefont {Tápai}}, \bibinfo {author}
  {\bibfnamefont {L.}~\bibnamefont {Singer}}, \bibinfo {author} {\bibfnamefont
  {S.}~\bibnamefont {Khan}}, \bibinfo {author} {\bibfnamefont {S.}~\bibnamefont
  {Fairhurst}}, \bibinfo {author} {\bibfnamefont {S.}~\bibnamefont {Kumar}},
  \bibinfo {author} {\bibfnamefont {A.}~\bibnamefont {Nielsen}}, \bibinfo
  {author} {\bibnamefont {SSingh087}}, \bibinfo {author} {\bibnamefont
  {shasvath}}, \bibinfo {author} {\bibfnamefont {I.}~\bibnamefont
  {Dorrington}},\ and\ \bibinfo {author} {\bibfnamefont {B.~U.~V.}\
  \bibnamefont {Gadre}},\ }\href {https://doi.org/10.5281/zenodo.3993665}
  {\bibinfo {title} {gwastro/pycbc: Pycbc release v1.16.9}} (\bibinfo {year}
  {2020})\BibitemShut {NoStop}%
\bibitem [{\citenamefont {{Biwer}}\ \emph {et~al.}(2019)\citenamefont
  {{Biwer}}, \citenamefont {{Capano}}, \citenamefont {{De}}, \citenamefont
  {{Cabero}}, \citenamefont {{Brown}}, \citenamefont {{Nitz}},\ and\
  \citenamefont {{Raymond}}}]{pycbc_inference_2019}%
  \BibitemOpen
  \bibfield  {author} {\bibinfo {author} {\bibfnamefont {C.~M.}\ \bibnamefont
  {{Biwer}}}, \bibinfo {author} {\bibfnamefont {C.~D.}\ \bibnamefont
  {{Capano}}}, \bibinfo {author} {\bibfnamefont {S.}~\bibnamefont {{De}}},
  \bibinfo {author} {\bibfnamefont {M.}~\bibnamefont {{Cabero}}}, \bibinfo
  {author} {\bibfnamefont {D.~A.}\ \bibnamefont {{Brown}}}, \bibinfo {author}
  {\bibfnamefont {A.~H.}\ \bibnamefont {{Nitz}}},\ and\ \bibinfo {author}
  {\bibfnamefont {V.}~\bibnamefont {{Raymond}}},\ }\href
  {https://doi.org/10.1088/1538-3873/aaef0b} {\bibfield  {journal} {\bibinfo
  {journal} {\pasp}\ }\textbf {\bibinfo {volume} {131}},\ \bibinfo {pages}
  {024503} (\bibinfo {year} {2019})},\ \Eprint
  {https://arxiv.org/abs/1807.10312} {arXiv:1807.10312 [astro-ph.IM]}
  \BibitemShut {NoStop}%
\bibitem [{\citenamefont {{Sathyaprakash}}\ and\ \citenamefont
  {{Schutz}}(2009)}]{sathyaprakash_schutz_2009}%
  \BibitemOpen
  \bibfield  {author} {\bibinfo {author} {\bibfnamefont {B.~S.}\ \bibnamefont
  {{Sathyaprakash}}}\ and\ \bibinfo {author} {\bibfnamefont {B.~F.}\
  \bibnamefont {{Schutz}}},\ }\href {https://doi.org/10.12942/lrr-2009-2}
  {\bibfield  {journal} {\bibinfo  {journal} {Living Reviews in Relativity}\
  }\textbf {\bibinfo {volume} {12}},\ \bibinfo {eid} {2} (\bibinfo {year}
  {2009})},\ \Eprint {https://arxiv.org/abs/0903.0338} {arXiv:0903.0338
  [gr-qc]} \BibitemShut {NoStop}%
\bibitem [{\citenamefont {{Kalaghatgi}}\ \emph {et~al.}(2020)\citenamefont
  {{Kalaghatgi}}, \citenamefont {{Hannam}},\ and\ \citenamefont
  {{Raymond}}}]{kalaghatgi_hannam_2020}%
  \BibitemOpen
  \bibfield  {author} {\bibinfo {author} {\bibfnamefont {C.}~\bibnamefont
  {{Kalaghatgi}}}, \bibinfo {author} {\bibfnamefont {M.}~\bibnamefont
  {{Hannam}}},\ and\ \bibinfo {author} {\bibfnamefont {V.}~\bibnamefont
  {{Raymond}}},\ }\href {https://doi.org/10.1103/PhysRevD.101.103004}
  {\bibfield  {journal} {\bibinfo  {journal} {\prd}\ }\textbf {\bibinfo
  {volume} {101}},\ \bibinfo {eid} {103004} (\bibinfo {year} {2020})},\ \Eprint
  {https://arxiv.org/abs/1909.10010} {arXiv:1909.10010 [gr-qc]} \BibitemShut
  {NoStop}%
\end{thebibliography}%
\end{document}